\documentclass[journal=jacsat,manuscript=article]{achemso}

\usepackage[version=3]{mhchem} 
\usepackage{color}
\usepackage{pbox}
\usepackage{chemfig}
\usepackage{amsmath}
\author{Erik  D. Hedeg{\aa}rd}
\affiliation[Lunds Universitet]
{Department of Theoretical Chemistry, Lund University, P.O. Box 124, SE-221 00, Lund, Sweden}
\email{erik.hedegard@teokem.lu.se}
\author{Ulf Ryde}
\affiliation[Lunds Universitet]
{Department of Theoretical Chemistry, Lund University, P.O. Box 124, SE-221 00, Lund, Sweden}
\email{ulf.ryde@teokem.lu.se}

\title[Molecular modeling of lytic polysaccharide monooxygenases]{Molecular mechanism of lytic polysaccharide monooxygenases}

\abbreviations{IR,NMR,UV}

\begin{document}

\begin{tocentry}
\end{tocentry}

\begin{abstract}
The lytic polysaccharide monooxygenases (LPMOs) are copper metalloenzymes that can enhance polysaccharide depolymerization through an oxidative mechanism and hence boost generation of biofuel from e.g.~cellulose.  By employing density functional theory in a  combination of quantum mechanics and molecular mechanics (QM/MM), we report the complete description of the molecular mechanism of LPMOs. The QM/MM scheme allows us to describe all  reaction steps with  a detailed protein environment and we show that this is necessary. 
  Several active species capable of abstracting a hydrogen from the substrate have been proposed previously and starting from recent crystallographic work on a substrate--LPMO complex, we investigate previously suggested paths as well as new ones. We  describe the generation of the reactive intermediates, the abstraction of a hydrogen atom from the polysaccharide substrate, as well as the final recombination step in which \ce{OH} is transferred back to the substrate. We show that a superoxo \ce{[CuO2]+} complex can be protonated by a nearby histidine residue (suggested by  recent mutagenesis studies and crystallographic work) and, provided an electron source is available, leads to formation of an oxyl-complex after cleavage of the O--O bond and dissociation of water. The oxyl complex either reacts with the substrate or is further protonated to a hydroxyl complex. Both the oxyl and hydroxyl complexes are also readily generated from a reaction with  \ce{H2O2}, which was recently suggested to be the true co-substrate, rather than \ce{O2}. The \ce{C-H} abstraction by the oxyl and hydroxy complexes is overall favorable with activation barriers of 69 and 94 kJ/mol, compared to the much higher barrier (156 kJ/mol) obtained for the copper--superoxo species. 
We obtain good structural agreement for intermediates for which structural data are available and the estimated reaction energies agree with experimental rate constants. 
Thus, our suggested mechanism is the most complete to date and concur with available experimental evidence. 
\end{abstract}

\section{Introduction}
Widespread and abundant polysaccharide bio-polymers constitute a major resource whose  utilization in production of biofuel or commercial chemicals would constitute a large step towards a more sustainable exploitation of resources.  
Unfortunately, this requires degradation of the polysaccharide into smaller sugars, which has shown to be a major obstacle and requires both hydrolytic enzymes and thermal work due to the remarkable stability of many naturally occuring polysaccharides.\cite{chang2007,himmel2007}    

A class of copper-dependent enzymes, called lytic polysaccharide monooxygenases (LPMOs), have been shown to enhance polysaccharide depolymerization, thereby providing a route to efficient conversion of polysaccharides into smaller carbohydrates.\cite{karkehabadi2008,harris2010,vaaje-kolstad2010,hemsworth2013a}. 
\begin{figure}    
\includegraphics[width=0.90\textwidth]{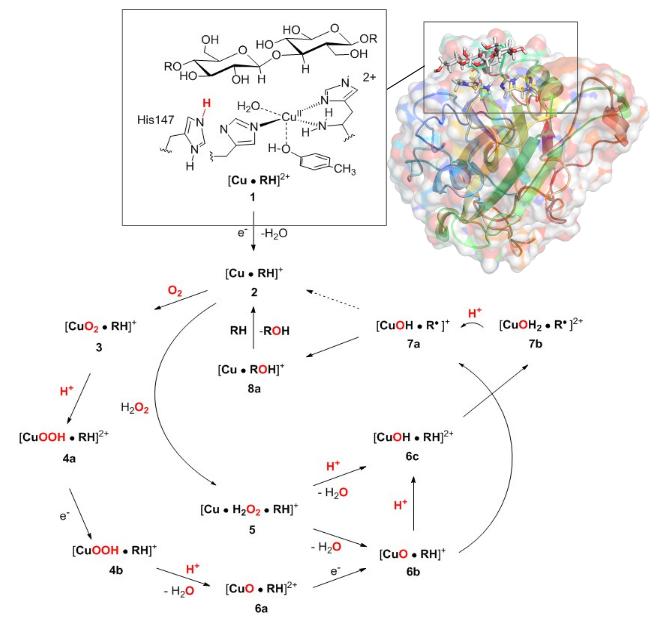} 
\caption{Fungal LPMO active site. Residue numbers refer to the enzyme from \textit{Lentinus similis}\cite{frandsen2016} (5ACF). 
\label{lpmo_active_site}}
\end{figure} 
The key to this enhancement is the ability of LPMOs to oxidise the \ce{C-H} bond of the glycoside linkage connecting the sugar units in polysaccharides, which  ultimately leads to cleavage of the glycoside link. A number of different LPMOs have been categorized, belonging to four distinct classes, AA9\cite{harris2010}, AA10\cite{quinlan2011}, A11\cite{hemsworth2014} and AA13.\cite{vu2014a,leggio2015}  
Speculations on  the underlying molecular mechanism have  begun\cite{beeson2015,span2015,walton2016,hemsworth2016,vaaje-kolstad2017}, but the picture is still far from complete. Mechanistic studies are complicated by the fact that 
the different LPMOs have remarkably varying amino-acid sequences and target a wide range of different polysacharride substrates,\cite{isaksen2014,bennati-Granier2015,agger2014,frommhagen2015} many of which are insoluble. Moreover, they have varying regioselectivity: some LPMOs oxidise only the C1 atom of the glycoside linkage, whereas others oxidise only the C4 atom, and still others can oxidise both C1 and C4. \cite{li2012,forsberg2014a,forsberg2014b,vu2014b,vaaje-kolstad2017}.  
However, a common feature of all LPMOs is the active site, in which a copper ion\cite{quinlan2011,phillips2011,beeson2012,horn2012,forsberg2011,hemsworth2013b,vaaje-kolstad2012,aachmann2012,vaaje-kolstad2013} is ligated by three nitrogen 
donor atoms in a so-called \textit{histidine brace} moiety, in which one histidine residue coordinates with the N$^{\epsilon 2}$ atom, whereas the other one (which is the amino-terminal residue) coordinates with both the side-chain N$^{\delta 1}$ and the backbone N atoms\cite{quinlan2011}.
A similar coordination environment is also seen in the particulate methane monooxygenases (but with an additional monodentate histidine ligand).\cite{solomon2014,cao2017} 

The net oxidation of a substrate \ce{RH} by \ce{O2} by the LPMOs proceeds under consumption of two electrons and two protons, as shown in Scheme \ref{net_reaction}. 
\begin{scheme}
\ce{RH + O2 + 2H+ + 2e- $\rightarrow$ ROH + H2O }
\caption{Reaction catalyzed by LPMOs. \label{net_reaction}}
\end{scheme}
The active site and some putative mechanisms for the reaction are shown in Figure \ref{lpmo_active_site}. In the resting state, the metal ion is in the Cu(II) state (\textbf{1}). We focus on AA9 LPMOs, for which the resting state without the substrate typically displays an octahedral coordination with an axial tyrosine ligand and two water molecules, one axial (\textit{trans} to tyrosine) and the other equatorial (\textit{trans} to the N-terminal amino group). When the polysaccharide substrate binds, the axial water molecule is probably displaced, as shown in a recent crystal structure of an LPMO--substrate complex\cite{frandsen2016}, although this does not seem to be universal for all substrates.\cite{simmons2017} Presumably, \textbf{1} is reduced to Cu(I) (\textbf{2}),  which leads to dissociation of the second, equatorial water molecule\cite{kjaergaard2014,frandsen2016} as is indicated for reaction \textbf{1}$\rightarrow$\textbf{2} in Figure \ref{lpmo_active_site}. 
 It is not known if this reduction takes place before or after the binding of the saccharide. The following steps are even more unclear. From \textbf{2}, most studies have suggested a mechanism employing \ce{O2} as co-substrate, leading to the oxygen species \ce{[CuO2]+} (\textbf{3}) and (after reduction, protonation and dissociation of water) \ce{[CuO]+} (\textbf{6b}) in Figure \ref{lpmo_active_site}. Both \textbf{3}\cite{phillips2011,beeson2012,li2012,beeson2015}  and \textbf{6b}\cite{beeson2015,lee2015,walton2016,hedegaard2017b,bertini2017} have been suggested as the reactive species that abstracts a hydrogen atom from the substrate. 
 
 Crystal structures of oxygen-bound LPMOs (without substrate) have been reported\cite{dell2017,bacik2017} and it is known that substrate-free LPMO can activate \ce{O2} and  produce \ce{H2O2}\cite{kittl2012,kjaergaard2014}.   
Meanwhile, studies on model systems have suggested both  hydroxyl\cite{dhar2015,walton2016} and hydroperoxyl complexes as reactive intermediates.\cite{neisen2017} 
 Quantum mechanical (QM) calculations on small cluster models of the enzyme have suggested that \ce{O2} is not reactive enough to abstract a hydrogen atom from the substrate, and that either oxyl\cite{kim2014,hedegaard2017b,bertini2017} or hydroxyl complexes\cite{hedegaard2017b} are more likely. By comparing calculated hydrogen bond-dissociation energies for small cluster models, we could recently show\cite{hedegaard2017b} that complexes with \ce{Cu-O} or \ce{Cu-OH} moieties (\textbf{6a}--\textbf{6c} in Figure \ref{lpmo_active_site}) are sufficiently reactive to abstract \ce{C-H}, whereas complexes with \ce{Cu-OO} or \ce{Cu-OOH} moieties (\textbf{3} and \textbf{4a}--\textbf{4b}) had too low bond-dissociation energies to abstract hydrogen. Yet, we did not consider the associated activation energies, how  complexes \textbf{6a}--\textbf{6c} may form or the effect of the surrounding protein, although we for complexes \textbf{1}--\textbf{3} have demonstrated that the protein imposes large strucutral changes on the active site\cite{hedegaard2017}. A recent study with larger QM-cluster models supported this and further confirmed that the hydrogen-abstraction by the oxyl-complex is favored over intermediates with intact \ce{O-O} bonds.\cite{bertini2017} Different  pathways for the formation of an oxyl-complex have been considered. Importantly, Bissaro et al.\cite{bissaro2017}  recently showed that \ce{H2O2}  rather than \ce{O2} could be the co-substrate.
 
 In this paper we have performed a full investigation of the reaction mechanism of the LPMOs.  The investigation is based on a recent crystal structure of a LPMO--oligosaccharide complex.\cite{frandsen2016} We investigate formation of both Cu--oxyl ($\textbf{6b}$)  and hydroxyl (\textbf{6c}) complexes along several suggested pathways, starting from the Cu(II)--superoxide complex (\textbf{3} in Figure \ref{lpmo_active_site}). Moreover, we investigate reaction paths where the Cu--oxyl and hydroxyl complexes are formed from hydrogen peroxide (\textbf{5}$\rightarrow$\textbf{6b} and  \textbf{5}$\rightarrow$\textbf{6c}). Several of the pathways discussed in this paper have not previously been considered, neither theoretically nor experimentally. However, the QM-cluster study in Ref.~\citenum{bertini2017} did consider coordination of \ce{H2O2} to Cu(I), while a very recent a QM/MM study\cite{wang2018} have also considered  generation of \textbf{6b} from peroxide. Unfortunately, the two studies  obtain contradictory results regarding  the \ce{H2O2} coordination to the Cu metal. We compare our results to these recent studies whenever possible.
 In all calculations we include the oligosaccharide substrate to investigate the reactivity of the reactive oxygen species in \ce{C-H} abstraction from the substrate. The calculations employ the combined QM and molecular mechancis approach (QM/MM) to include possible effects of the surrounding protein.   
 

\section{Models and methods}

\subsection{General computational setups}

Calculations in this paper were performed with the QM/MM approach, using the  QM software  
Turbomole 7.1\cite{dm891013abhhk} and the MM software AMBER 14\cite{amber14}. 
The QM/MM calculations were performed with the {\sc ComQum} interface\cite{ryde1996,ryde2001}, which combines these two programs.  
In {\sc ComQum}, the total studied system is divided into three subsystems, denoted systems 1, 2 and 3. System 1 is described with a QM method, while systems 2 and 3 both are described with an MM force field. The structure of system 2 can optionally be optimised at the MM level, whereas system 3 is always kept fixed at the starting structure.   
When there is a bond between systems 1 and 2 (a junction), the hydrogen link-atom approach is employed: the QM region is capped with hydrogen atoms (hydrogen link atoms), 
the positions of which are linearly related to those of the corresponding carbon atoms (carbon link atoms) in the full system.\cite{ryde1996,reuter2000} 
 
The total QM/MM energy is calculated as 
\begin{equation}
 E_{\text{QM/MM}} = E_{\text{QM+ptch}} + E_{\text{MM123}} - E_{\text{MM1}} . \label{QQ_MM_opt}
\end{equation}
$E_{\text{QM+ptch}}$ is the QM energy of system 1, including hydrogen link atoms and a point-charge model of systems 2 and 3 
(taken from the Amber force field and excluding only the carbon link atoms).\cite{hu2011} $E_{\text{MM123}}$ is the total MM energy of the full 
system (but with the charges of the QM system zeroed) and $E_{\text{MM1}}$ is the MM energy of system 1 (still with zeroed charges; it is included to avoid double-counting of the energy of system 1). We will for some reactions discuss the electrostatic effect of the protein,  which is calculated as 
\begin{align}
E_{\text{ptch}} = E_{\text{QM+ptch}} - E_{\text{QM}}\label{E_ptch}
\end{align}   
where $E_{\text{QM}}$ is taken from a calculation in vacuum with the QM/MM optimised structure. 

All QM calculations employed density functional theory and reported energies were obtained with the dispersion-corrected TPSS-D3 functional\cite{tao2003,grimme2010}, employing a def2-TZVPP basis set\cite{bs920815sha} (unless otherwise specified). The energies were also checked (using the same structures and basis set) with the B3LYP-D3\cite{becke1988,becke1993,lee1988} functional. All energies were obtained as single-point calculations on structures optimised with TPSS-D3 and the def2-SV(P) basis set\cite{bs920815sha,eichkorn1997}.
  In general, we only report QM/MM energies, but a more detailed breakdown of $E_{\text{QM/MM}}$ in the QM and MM energy components (and $E_{\text{ptch}}$ from Eq.~\ref{E_ptch}) is provided in the supporting information (SI). The SI also contains a more detailed account of the computational and protein setup (including the alternate configurations and protonation states of individual amino acids). 
  
  The QM system (system 1) consisted of the copper ion and its first coordination sphere. For all intermediates this is the imidazole ring of His78 and the phenol ring of Tyr164, both 
capped with a hydrogen atom replacing C$^{\alpha}$. The entire His1 residue, which coordinates to 
Cu through the terminal amino group, as well as the imidazole side chain, was alsoincluded. The neighboring Thr2 residue was included up to the C$^{\alpha}$ atom, which was replaced by a hydrogen atom.  
The last ligand in the first coordination sphere varies between the different intermediates in Figure \ref{lpmo_active_site}. In \textbf{1} and \textbf{2}, the ligand is \ce{H2O}. In the next steps the water ligand is replaced in accordance with the various steps shown in Figure \ref{lpmo_active_site}.
In addition, the two first glucose rings of the substrate were also included in system 1, whereas the third glucose unit was always described by MM.  Apart from a few initial calculations, the His147 residue was generally also included (both in HID, HIE and HIP forms). Representative examples of the QM systems employed are shown in Figure S1 in the SI.

\section{Results \label{results}} 
In the first part of this section, we compare the obtained QM/MM structures with structures from earlier QM-cluster\cite{kim2014,gudmundsson2014} and QM/MM studies\cite{hedegaard2017}. Here we focus on intermediates \textbf{1}--\textbf{3}, for which we briefly discuss differences in our structures compared to previous theoretical work.   
Next, we discuss the formation of the reactive oxygen species, i.e.~\textbf{6a}--\textbf{6c} 
in Figure \ref{lpmo_active_site}, before we finally investigate the \ce{C4-H} abstraction by \textbf{6b} and \textbf{6c} (reactions \textbf{6}$\rightarrow$\textbf{7}), as well as the recombination to form a \ce{Cu+} species and \ce{ROH} (reactions \textbf{7}$\rightarrow$\textbf{8}).


\subsection{The resting state and the initial reduction}

We start by qualitative discussion of the structural changes accompanying the first reduction of the \ce{[Cu(H2O)]^{2+}} state  (reaction \textbf{1}$\rightarrow$\textbf{2} in Figure \ref{lpmo_active_site}). A comparison between our previous QM/MM results, QM-cluster calculations and experimental results were given in Ref.~\citenum{hedegaard2017}, and we here focus on a comparison of differences for the \textbf{1} and \textbf{2} states to our previous QM/MM structures without substrate\cite{hedegaard2017}. We note that this comparison cannot directly be carried out with available experimental data, since the LPMO--substrate crystal structure  has \ce{Cl-} coordinating to Cu in the equatorial position and an empty axial coordination site. In our current QM/MM setup, the axial coordination site is kept empty (unlike our previous QM/MM structures), yielding a five-coordinate copper site.     

\begin{figure}[tbh!]
\centering
  \includegraphics[scale=0.30]{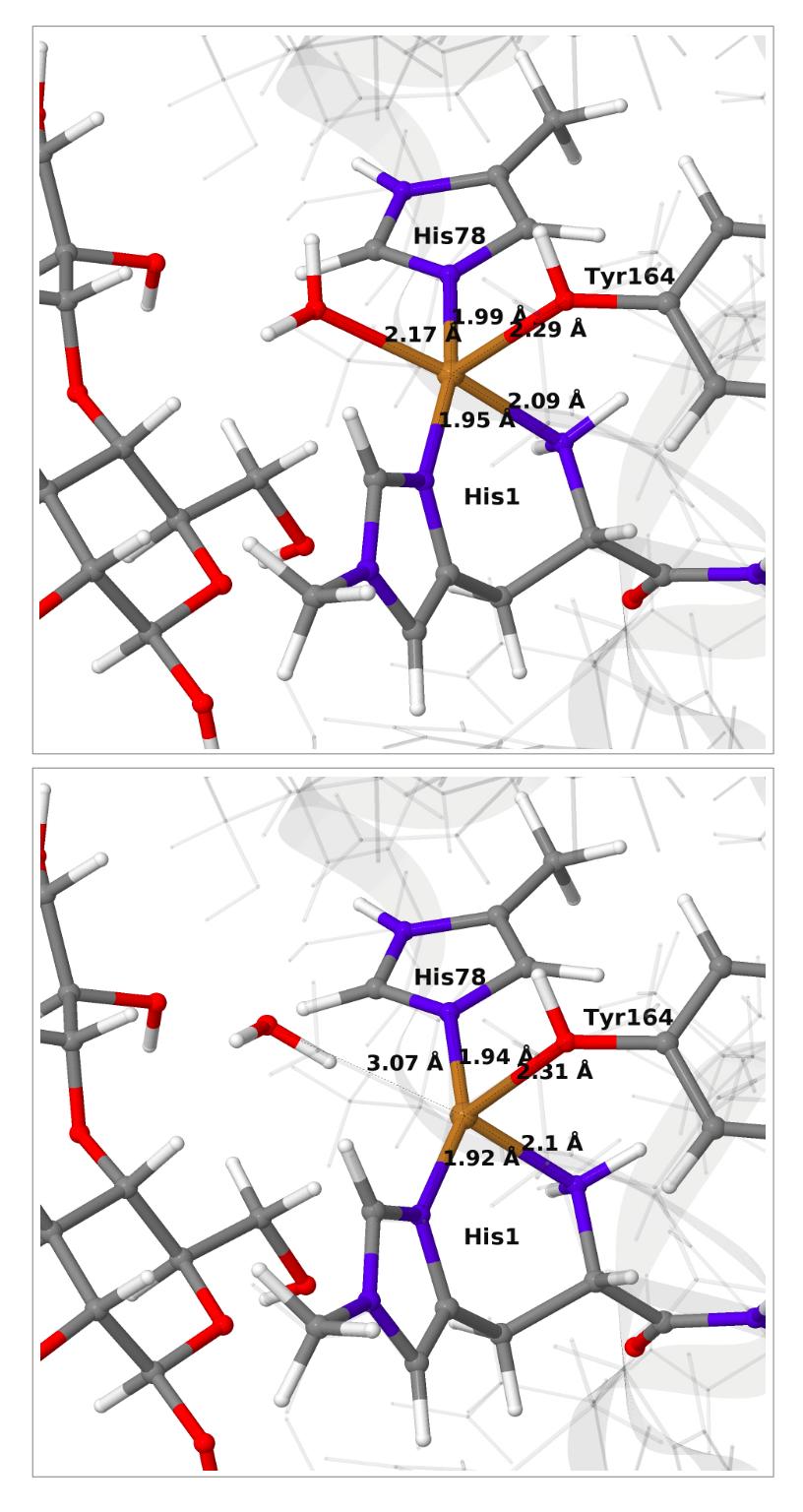}
 \caption{QM/MM optimised \ce{[Cu(H2O)]^{2+}} (\textbf{1}, \textit{upper}) and \ce{[Cu(H2O)]^{+}} (\textbf{2}, \textit{lower}) states. The optimisations were carried out with TPSS-D3/def2-SV(P) and system 2 fixed.    \label{cu-rest_state_and_cu-rest_state_red} }  
\end{figure}

Experimental and computational studies for LPMOs of both the AA9 and AA10 families have shown that the reduction (\textbf{1}$\rightarrow$\textbf{2}) is accompanied by dissociation of one or both water molecules from the copper ion.\cite{gudmundsson2014,kim2014}
Here, we have denoted the reduced state \ce{[Cu(H2O)]+} (\textbf{2}), to emphasize that the water molecule always is included in the QM system for  \textbf{1} and \textbf{2}, although it might not bind directly to Cu(I). For the AA9 enzymes, the reduction is also associated with an elongation of the \ce{Cu-O} bond to the axial tyrosine\cite{gregory2016,gudmundsson2014,kim2014,kjaergaard2014,hedegaard2017}. 
This Tyr ligand is typically not present in the AA10 LPMOs, which results in a different (trigonal bipyramidal) geometry in the AA10 LPMOs\cite{hemsworth2013b,gregory2016}. 
The active sites in our QM/MM optimised structures are shown in Figure \ref{cu-rest_state_and_cu-rest_state_red}. The Cu--O distance to the water ligand in \textbf{1} is 2.17 {\AA} and when one electron is added to this structure, the water molecule dissociates to form \textbf{2} (the Cu--O distance is 3.07 \AA, as can be seen in Figure \ref{cu-rest_state_and_cu-rest_state_red}). 
 
 In contrast to our previous QM/MM calculations without substrate\cite{hedegaard2017}, the \ce{Cu-O} bond length to tyrosine is remarkably constant (2.29 and 2.31 \AA, respectively). It has been argued that this bond shortens upon substrate binding\cite{frandsen2016}. Our results suggest the bond may remain short also in the \textbf{2} state due to the lack of the axially coordinated water in the substrate--LPMO complex\cite{frandsen2016}.
 We note that an elongation of this bond was suggested in the QM-cluster results of Ref.~\citenum{bertini2017}. We have previously discussed this bond\cite{hedegaard2017}, which has shown to be flexible and rather sensitive to the computational setup. We investigated wheather the different starting structures in Ref.~\citenum{bertini2017} and our study could be the underlying reason. However, the overlay of the structures employed here (5ACF\cite{frandsen2016}) and by Bertini et al.\cite{bertini2017} (4EIS\cite{li2012}) in Figure S2 shows that they are fairly similar. It seems therefore more likely that slight differences in basis set and functional choice, or possibly the lack of dispersion corrections in Ref.~\citenum{bertini2017}, is the underlying reason for this discrepancy.
 
 To ensure that the employed QM/MM model is sufficiently accurate, we also carried out optimisations with system 2 optimised with MM. As can be seen in Figure S3 in the SI, nearly identical conclusions were obtained and the distances within the first ligand-sphere change little when system 2 is relaxed. However, some changes in the second coordination sphere are observed, shifting the position of the active site (cf.~Figure S4). This indicates that an energetically accurate description may require a larger QM region for reaction \textbf{1}$\rightarrow$\textbf{2}, underlining that our discussion of this step is qualitative.  It is nevertheless reassuring that our model can reproduce the experimentally observed water dissociation upon reduction. As will be discussed in the next subsection, the structure of the \ce{[CuO2]+} (\textbf{3}) state changes much less when system 2 is relaxed. Before discussing  this state, we note that we also optimised the intermediates \textbf{1} and \textbf{2} with the larger def2-TZVPD basis, which lead to only minor changes.

\subsection{The superoxide state} \label{super-oxide-state}

We next consider the \ce{[CuO2]^{+}} state (\textbf{3} in Figure \ref{lpmo_active_site}), i.e. the state formed by binding of \ce{O2} to the reduced active site. As for \textbf{1} and \textbf{2}, we have previously investigated \textbf{3}\cite{hedegaard2017}, but for a different LPMO and without any substrate. It should be noted that it is not known wheather this state is formed before or after binding of the substrate, since all available crystal structures of oxygen-bound states are substrate-free and their reduction level is unknown.\cite{dell2017,bacik2017}  

\begin{figure}[tbh!]
\centering
  \includegraphics[scale=0.30]{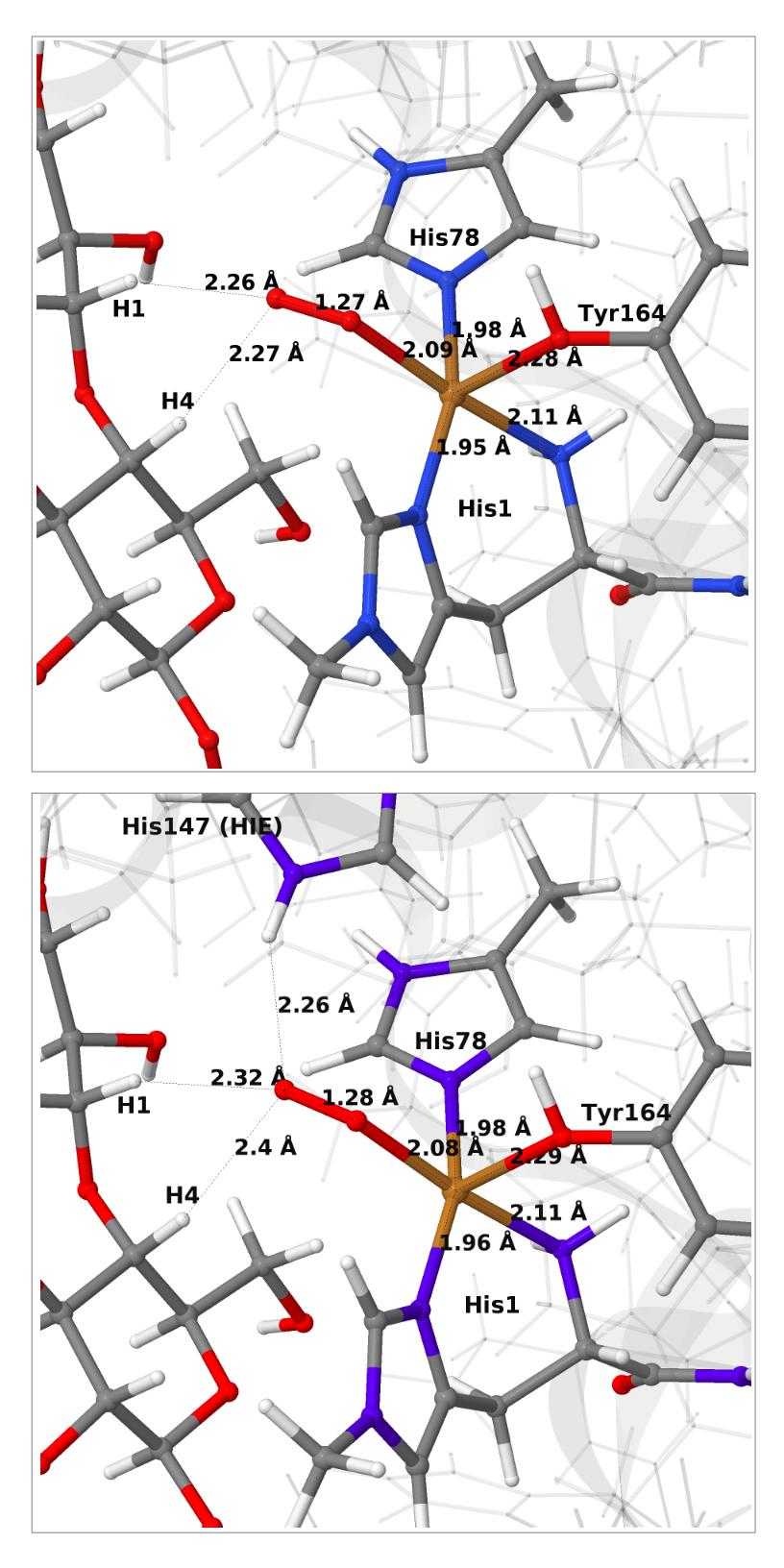}
 \caption{QM/MM optimised \ce{[CuO2]^{+}} (\textbf{3}) without (\textit{upper}) and with (\textit{lower}) His147 included in the QM region. Both optimisations were carried out with TPSS-D3/def2-SV(P) and system 2 fixed. \label{cu_o2_qm_structures} }  
\end{figure}

The optimised structure for \textbf{3} are shown in Figure \ref{cu_o2_qm_structures} (upper part); selected structural parameters are compared to earlier results in Table S1 in the SI. The optimised structure has a relative short \ce{Cu-O} Tyr bond (2.28~\AA), 
compared to previous QM/MM\cite{hedegaard2017} (2.89~\AA) and 
QM-cluster\cite{kjaergaard2014,bertini2017} results (3.4--4.1~\AA). Again, the main difference between our QM/MM results with and without substrate is a water molecule coordinated in the axial position in the previous structures. Hence, also for \textbf{3} we ascribe the shorter \ce{Cu-O} Tyr bond to the lack of this water molecule. Again, the result in Ref.~\citenum{bertini2017} stands out with a very  long Cu--O distance for the tyrosine ligand (4.1~\AA), despite the lack of an axial water molecule. The experimental structure of the LPMO--substrate complex\cite{frandsen2016} had an \ce{Cl-} ion coordinating in the equatorial position (where \ce{O2-} coordinates in Figure \ref{cu_o2_qm_structures}). The \ce{Cl-} ion is known as a \ce{O2-} mimic and the experimental \ce{Cu-O} Tyr bond of 2.5 {\AA} also shows a decrease, compared to the recent substrate-free structure of \ce{O2} bound intermediate,\cite{dell2017} in which the corresponding distance is between 2.6 and 2.7 {\AA}. 
The experimentally observed decrease is thus 0.1--0.2 {\AA}. In our optimized structures of \textbf{3} with and without substrate (cf.~Table S1), we obtain a decrease around 0.1 {\AA}, which fits well to the experiment.  

Allowing system 2 to relax for \textbf{3} has a much smaller effect than for intermediates \textbf{1} and \textbf{2}. The optimised structures are compared in Figure S5 in the SI.  Both the Cu--ligand distances and the overall placement of the active site are almost identical for the structures with and without system 2 relaxed. The same is true when employing a larger basis set (def2-TZVPD), as was also seen for the substrate-free QM/MM results \cite{hedegaard2017} (cf.~entries 3, 6 and 7 in Table S1). 
  
The QM calculations indicate that the \ce{[CuO2]+} moiety consists of Cu(II) and a superoxide ion, each with one unpaired electron. These two electrons can have either parallel or antiparallel spin, giving rise to triplet or (open-shell) singlet states. The two spin states have nearly identical structures (selected distances 
are reported in the lower part of Table S1). We find that the triplet is 12--14 kJ/mol more stable than the open-shell singlet state. 
Spin-state splitting energies are reported in Table S2 and they
are similar to what was obtained for the substrate-free LPMO in Ref.~\citenum{hedegaard2017}. We therefore focus on the triplet state in the following.

Next, we included the second-sphere residue His147 in the QM system in various protonation states. This residue has been suggested to stabilize the \ce{O2-} intermediate\cite{dell2017} and it is also a putative proton donor.\cite{dell2017} The latter implies that it is in the doubly protonated state (denoted HIP in the following), which is also the most probable form at the pH value of maximum activity for many cellulases (5.0). It has also been shown that the activity of some LPMOs is reduced\cite{span2017} by mutations of His147, suggesting a role of His147 in the mechanism. The role could be that His147 facilitates the formation of \textbf{4a} and \textbf{4b}, and perhaps ultimately the Cu--oxyl (\textbf{6a}--\textbf{6b}) or hydroxy (\textbf{6c}) species. We investigate the first step in this section, \textbf{3}$\rightarrow$\textbf{4a}, and several possibilities for the fate of \textbf{4a} below.  

\begin{figure}[tbh!]
\centering
  \includegraphics[scale=0.25]{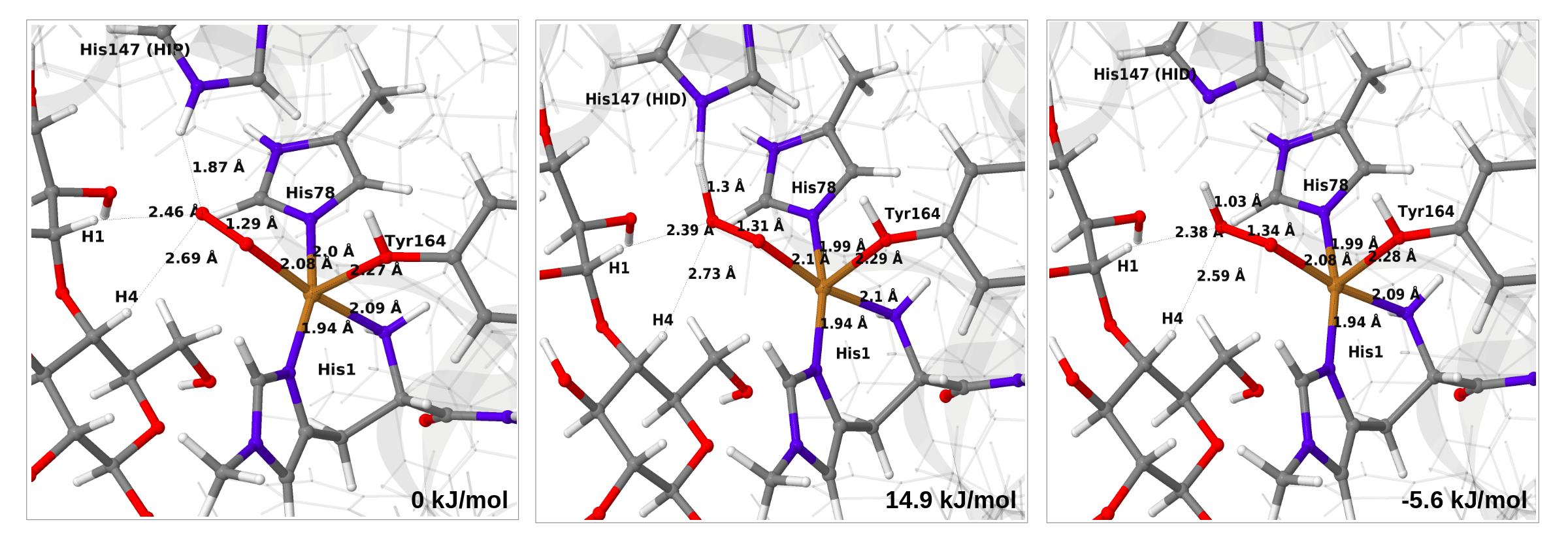}
 \caption{Reactant \textbf{3} (\textit{left}), transition state (\textit{middle}) and product \textbf{4a} (\textit{right}) for the reaction \textbf{3}$\rightarrow$\textbf{4a}. All structures were optimised with TPSS/def2-SV(P) and system 2 fixed. \label{cu_o2_his147_h_transfer} }  
\end{figure}

The optimised structure of \textbf{3} including His147 in the HIE form (i.e. protonated only on the N$^{\epsilon 2}$ atom) is shown in Figure \ref{cu_o2_qm_structures} (lower part), whereas we show the corresponidng complex with His147 in the HIP form in Figure \ref{cu_o2_his147_h_transfer} (left).     
The inclusion of His147 in the QM region has only a marginal effect on the Cu--ligand distances. The effect is somewhat larger when considering the distances between the \ce{O2-} ion and the substrate, which are 2.26--2.27 {\AA} for \textbf{3}, but increase to 2.32--2.40 {\AA} for the HIE form and 2.46--2.69 {\AA} for the HIP form. Another significant change between the HIE and HIP variants is that the distance between the \ce{H^{$\epsilon$2}} proton and \ce{O2-} changes from 2.26 to 1.87 {\AA}, reflecting the positive charge in the latter structure. This change most likely facilitates proton transfer.  

The structures along the reaction path for such a transfer are shown in Figure \ref{cu_o2_his147_h_transfer}, along with the calculated QM/MM energies (see further Table S3 and Figure S6; the latter shows energies for individual \ce{O-H} distances). It can be seen that the reaction is nearly thermoneutral with a small barrier (15--18 kJ/mol, depending on the functional), showing that protonation is facile.
 As can be seen from the structures in Figure \ref{cu_o2_his147_h_transfer}, the reaction proceeds with small changes in the structures, even for distances within the first ligand sphere. 
Interestingly, the electrostatic contributions from the environment increase steady as the proton moves from His147 to \ce{O2}. This is shown in Figure S6 and it can be seen that in vacuum the reaction is uphill by more than 50 kJ/mol (cf.~$\Delta E_{\text{QM}}$ in Table S3). 

 The net reaction (Scheme \ref{net_reaction}) involves addition of two electrons and two protons, but the order of these additions is unknown. Although the reaction might proceed with reduction first or with simultaneous reduction and protonation, the results in this sections shows that it is reasonable to assume that the first step in the reaction chain is the protonation of \textbf{3} to \textbf{4a}, as this reaction was found to be facile.  
The further reactions from \textbf{4a} to form oxyl or hydroxyl intermediates (\textbf{6} in Figure \ref{lpmo_active_site}) will be investigated below. 

\subsection{Formation of Cu--oxyl and hydroxyl species from \ce{O2} \label{protonation}}

 We now proceed to consider formation of Cu--oxyl or hydroxyl species with \ce{O2} as substrate, i.e., by subsequent protonation and reductions as laid out in the route \textbf{3}$\rightarrow$\textbf{4}$\rightarrow$\textbf{6} in Figure \ref{lpmo_active_site}. We consider several possible pathways to generate \ce{[CuO]+} (\textbf{6b}) from  \ce{[CuOOH]^{2+}} (\textbf{4a}). 
The proton donor is always the His147 residue (in the HIP form).   
 
 Direct cleavage of the O--O bond and dissociation of water directly from \textbf{4a} (without any reduction) is uphill by 76 kJ/mol and requires a high activation barrier of 111 kJ/mol (with the TPSS functional).
 This is quite expected as it gives rise to a \ce{[Cu-O]^{3+}} species. Additionally, we also investigated a pathway in which \ce{O-O} is cleaved after reduction of \textbf{4b}, which would mean \ce{O-O} cleavage occurs after two-electron reduction. However, this also turned out to be unfavorable, since  it was found the reduction of \textbf{4b} did not lead to reduction of the \ce{[CuOOH]^{+}} unit, but rather reduced histidine (His147). A similar conclusion for this state was obtained in our previous QM-cluster study.\cite{hedegaard2017b}  
 
\begin{figure}[tbh!]
\centering
  \includegraphics[scale=0.265]{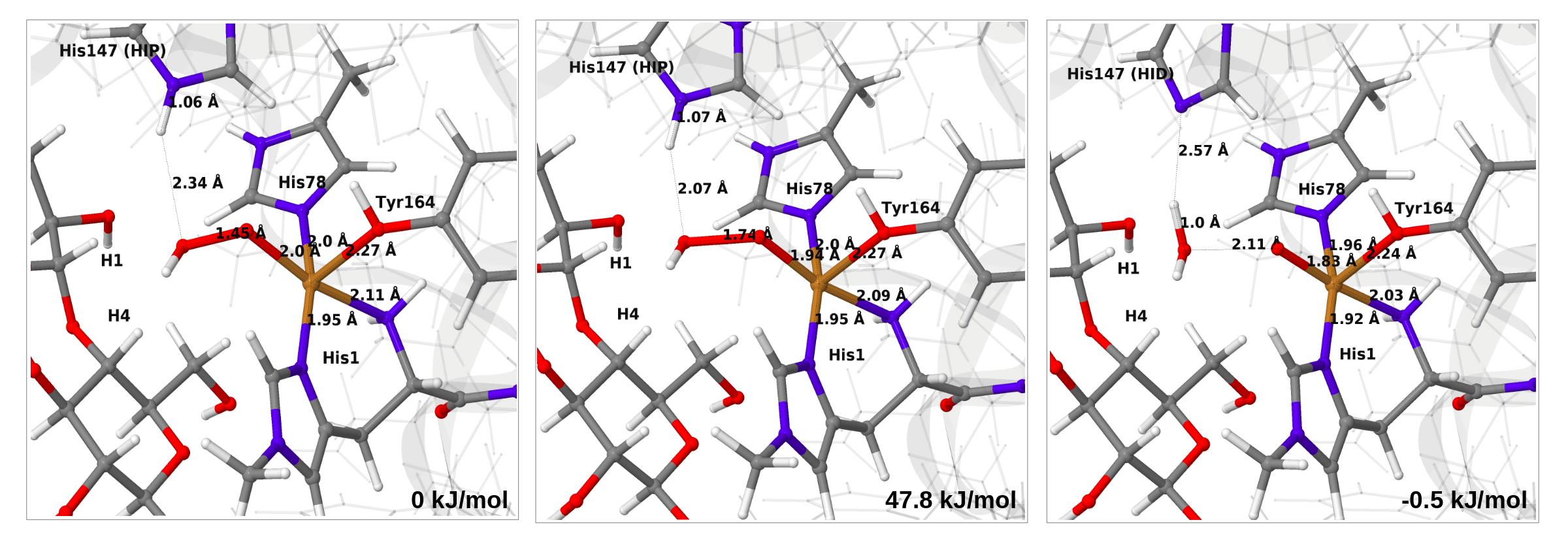}
 \caption{Reactant \textbf{4b} (\textit{left}), transition state (\textit{middle}) and product \textbf{6a} (\textit{right}) for reaction \textbf{4b}$\rightarrow$\textbf{6a}. Structures were optimised with TPSS/def2-SV(P) and energies were obtained with TPSS-D3/def2-TZVPP. \label{cu_ooh_red_his147_h2o_diss} }  
\end{figure}

 Therefore, we focus on the one-electron reduced state and the \textbf{4b}$\rightarrow$\textbf{6a} reaction. 
QM/MM energetics, structures and selected bond lengths for the reactant, transition state and product are shown in Figure \ref{cu_ooh_red_his147_h2o_diss}. The main difference between reactant and product is the short \ce{Cu-O} bond (1.8 \AA) in the oxyl species (\textbf{6a}) compared to the parent \textbf{4b}, for which the \ce{Cu-O} bond is 2.0 \AA. 
    The calculated activation barrier is 48 kJ/mol and the reaction is thermoneutral with the TPSS functional. Results with the B3LYP functional, together with the QM/MM energy components are provided in Table S4. Interestingly, they show that the reaction energy strongly depend on the employed functional: With B3LYP, the reaction is uphill with 54 kJ/mol. The activation energy is less functional dependent, with a B3LYP value of 66 kJ/mol. Still, the reaction is predicted to be feasible both with TPSS and B3LYP, although in the latter case the product, \textbf{6a}, is expected to have a shorter life time.  

Further analysis of the reaction shows that the electrostatic contributions increase steadily with increased \ce{O-O} distance ($E_{\text{ptch}}$ is shown as a function of the O--O distance in Figure S8). The effect is significant already for the transition states: depending on the functional, the effect is 50--100 kJ/mol for the activation energy and around 100 kJ/mol for the reaction energy (cf.~Table S4).  Thus, reaction \textbf{4b}$\rightarrow$\textbf{6a} is much less favorable in vacuum (by 96--102 kJ/mol), compared to the result from $E_{\text{QM+ptch}}$. Accordingly, QM-cluster calculations with small clusters may fail to reproduce the reaction profile, emphasizing the importance of including the protein environment. We have carefully analyzed the underlying wave functions from QM/MM and QM calculations, showing that the large differences between QM and QM/MM is not a result of   different electronic states. Rather, it is the change of protonation state for His147 (from HIP to HID) that gives a large environment effect.

\subsection{Formation of Cu--oxyl and hydroxy species from \ce{H2O2}}

In light of the recent proposal that \ce{H2O2} is the co-substrate, we have also investigated coordination of \ce{H2O2} to the Cu(I) ion, i.e. species \textbf{5} in Figure \ref{lpmo_active_site}.  Initially, we investigated all three forms of His147, namely HIP, HIE and HID (i.e. protonated only on the N$^{\delta 1}$ atom) in the singlet spin state. However, the HID form was found to be less stable than HIE and we therefore focus on the HIE and HIP forms (we also investigated the corresponding triplet spin-states which, as expected, always was higher in energy than the singlet).  

\begin{figure}[tbh!]
\centering
\includegraphics[width=0.35\textwidth]{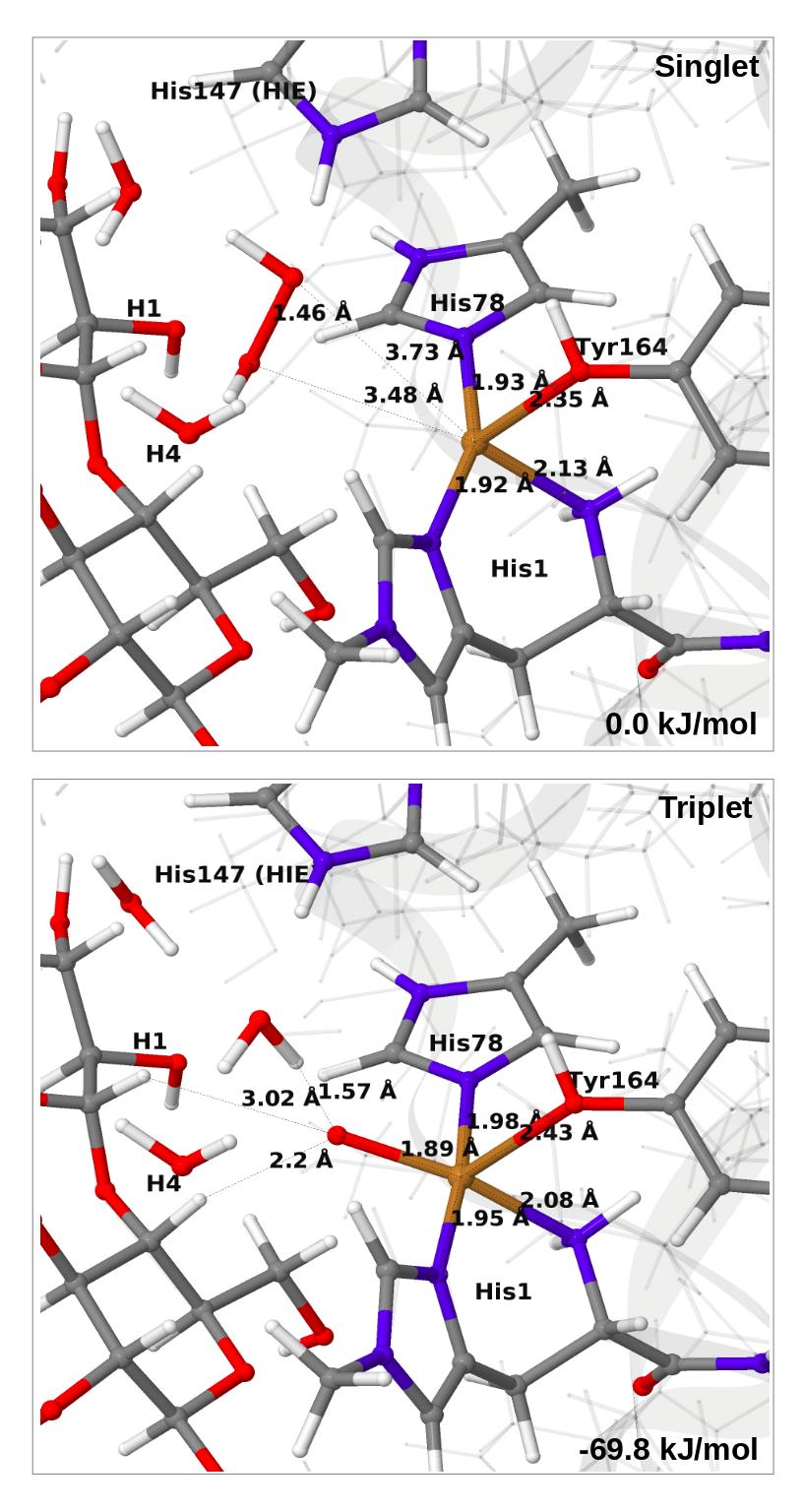}  
\caption{Optimised structures of intermediates \textbf{5} (\textit{upper}) and \textbf{6b} (\textit{lower}) in the HIE147 state. Structures were optimised with TPSS/def2-SV(P) and energies were obtained with TPSS-D3/def2-TZVPP. \label{cu-ohoh_hie147}}
\end{figure}

For all three protonation states of His147, coordination of \ce{H2O2} turned out to be unstable, which is not unexpected, considering that Cu(I) typically prefers low coordination numbers. For the HIE variant, we have optimised a second-sphere singlet state (\textbf{5}) shown in Figure \ref{cu-ohoh_hie147} (\textit{upper} part) with  \ce{Cu-O} distances of 3.5--3.7 \AA. A similar second-sphere complex of Cu(I) and \ce{H2O2} was described in Ref.~\citenum{wang2018}, both for a small QM-cluster model and with QM/MM, although the Cu--\ce{H2O2} distances were somewhat smaller in their QM/MM structure (around 2.8 {\AA}). In our case, the position of \ce{H2O2} is over the Cu(I) ion, interacting with the H$^{\epsilon 2}$ proton of HIE147 and also with Gln162, as was also found in Ref.~\citenum{wang2018}. In addition, \ce{H2O2}  interacts with two nearby water molecules and these were included in the QM region for all states with HIE147. Notably, the crystallographic study in Ref.~\citenum{dell2017} found that the site occupied by \ce{H2O2} in Figure \ref{cu-ohoh_hie147} can be occupied by \ce{O2} in what they called a pre-bound state. 
 
 Attempts to coordinate \ce{H2O2} to the Cu(I) ion lead to structures in which the O--O bonds are very long with \ce{O-O} distances of 2.0 (singlet) and  2.1 {\AA} (triplet), respectively (see Figure S7 in the SI). These intermediates are similar to the "caged \ce{OH$^{\bullet}$}" structures  obtained in Ref.~\citenum{wang2018}. However, they were appreciably less stable than a structure in which \ce{H2O2} is cleaved to \ce{H2O} and \ce{[CuO]+} (\textbf{6b}) in the triplet state (with His147 in the HIE state). This intermediate formed spontaneously under the attempt to optimise \textbf{5} with \ce{H2O2} coordinating to Cu(I) in the triplet state. The resulting structure is shown in Figure \ref{cu-ohoh_hie147} (\textit{lower}). It is about 70 kJ/mol more stable than \textbf{5} (in the singlet state). Interestingly, the \ce{C4-H} distance is 2.2 {\AA}, so it is appearently set up for hydrogen abstraction. 

\begin{figure}[tbh!]
\centering
\includegraphics[width=0.35\textwidth]{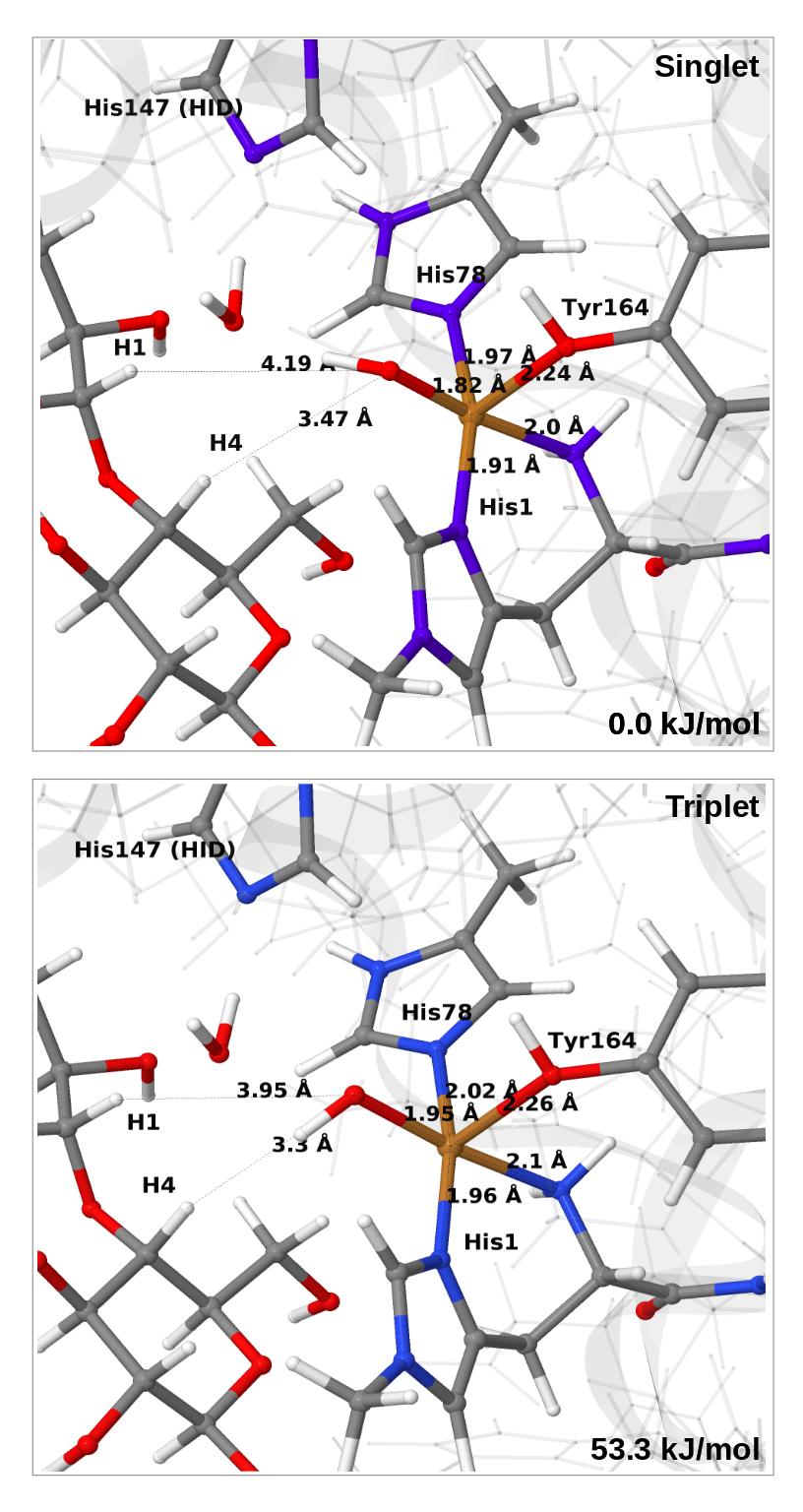}  
\caption{Optimised structures of intermediate \textbf{6c} in the singlet (\textit{upper}) and triplet states (\textit{lower}) with HIP147. Structures were optimised with TPSS/def2-SV(P) and energies were obtained with TPSS-D3/def2-TZVPP. \label{cu-ohoh_hip147}}
\end{figure}

When His147 is protonated (HIP), we obtain only states in which the \ce{O-O} bond is cleaved, i.e. \textbf{6c} is formed  spontaneously, independent of the spin-state and also when starting from structures with \ce{H2O2} pointing away from HIP147. His147 was kept in the HIE form in Ref.~\citenum{wang2018}, and direct comparisons are therefore not possible. 
 The various structures and the relative energies between the singlet and triplet are shown in Figure \ref{cu-ohoh_hip147}. The singlet is in this case favored by 53 kJ/mol. The differences between the two spin states for the bond distances within the first coordination sphere are small, whereas the distance to \ce{C4-H} is 3.3--3.5 {\AA}, i.e. slightly longer than for \textbf{6b}, but still allowing a facile \ce{C-H} transfer (as discussed further below). Thus, the results show that \ce{H2O2} is a possible co-substrate, as both \textbf{6b} and \textbf{6c} are readily generated from \ce{H2O2}. Our calculations predict that \ce{H2O2} reacts by a second-sphere mechanism, in which His147 positions the co-substrate in proximity to Cu(I) before the cleavage of the O--O bond.

\subsection{\ce{C-H} activation reaction}

Next, we investigate abstraction of the \ce{C4-H} atom from the polysacharide substrate. For the \ce{[CuO2]+} (\textbf{3}) intermediate, the reaction energy is 140 kJ/mol and the activation barrier is more than 150 kJ/mol (see Table S5 and Figure S9), making this reaction highly unlikely. We recently found the same employing a QM-cluster approach, although no activation barriers were calculated\cite{hedegaard2017b}. Another recent QM-cluster study also came to the same conclusion.\cite{bertini2017} Therefore, we focus on the \ce{[CuO]+} (\textbf{6b}) and \ce{[CuOH]^{2+}} (\textbf{6c}) intermediates in the following.
 However, before investigating the reactions \textbf{6b}$\rightarrow$\textbf{7a} and \textbf{6c}$\rightarrow$\textbf{7b}, we investigated the protonation state of His147 for the reactants and products. For intermediates \textbf{6b} and \textbf{7a}, the HIE state is 120 and 122 kJ/mol more stable than the HID state, respectively. The corresponding energy differences for \textbf{6c} and \textbf{7b} are 116 kJ/mol and 187 kJ/mol, also in favor of the HIE state. On the one-hand, this large energy-difference in favor of the HIE tautomer suggest that the abstraction reaction occurs from the HIE form, unless the barrier is significantly higher in this form. 
 On the other hand, the HID form is more natural to employ for the \ce{C4-H} abstraction, as this form is obtained directly from reactions \textbf{5}$\rightarrow$\textbf{6c} and \textbf{4}$\rightarrow$\textbf{6} (in which HIP donates a proton and thereby becomes HID). With these considerations in mind, we decided to investigate the \ce{C4-H} abstraction with both the HIE and HID forms of His147.
It turned out, that for \textbf{6b} and to a lesser degree \textbf{6c}, the protonation state of the nearby His147 residue strongly influences the activation energies. 

\begin{figure}[tbh!]
\centering
   \includegraphics[scale=0.27]{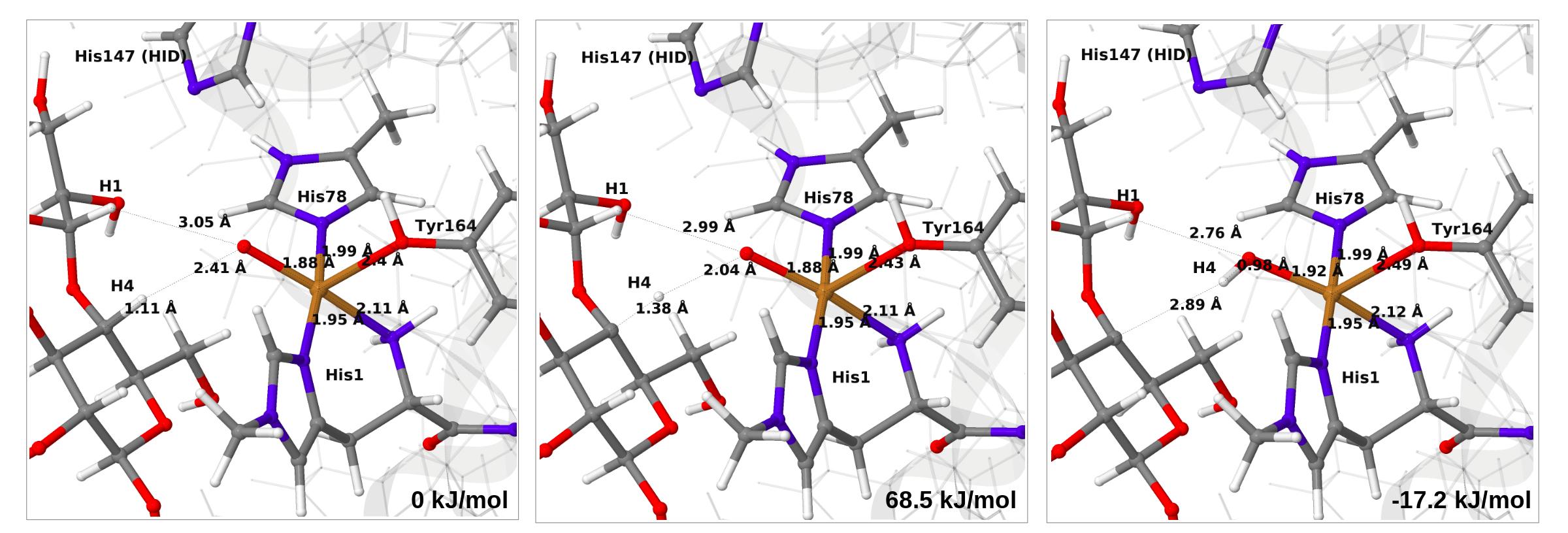}
 \caption{Reactant (\textbf{6b}, \textit{left}), transition state (\textit{middle}) and product (\textbf{7a}, \textit{right}) for the \textbf{6b}$\rightarrow$\textbf{7a} reaction with His147 in the HID state. Structures were optimised with TPSS/def2-SV(P) and energies were obtained with TPSS-D3/def2-TZVPP. \label{cu-o_his_h-transfer_hid}}  
\end{figure}

\begin{figure}[tbh!]
\centering
   \includegraphics[scale=0.27]{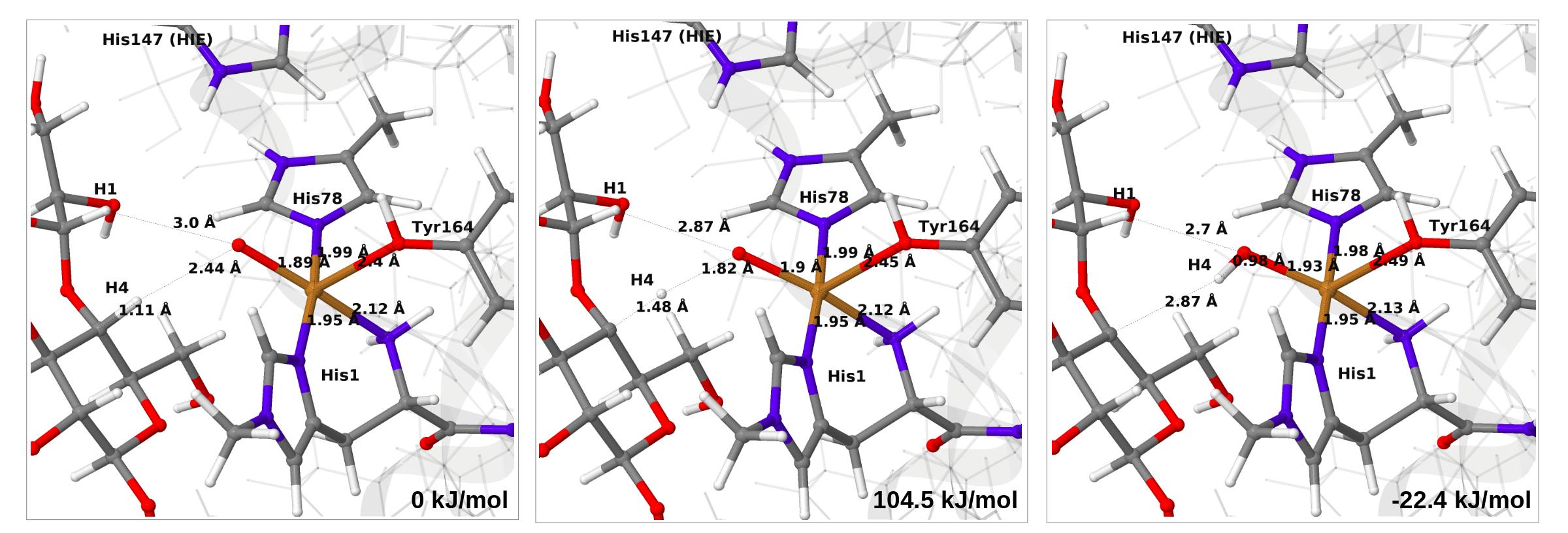}
 \caption{Reactant (\textbf{6b}, \textit{left}), transition state (\textit{middle}) and product (\textbf{7a}, \textit{right}) for the \textbf{6b}$\rightarrow$\textbf{7a} reaction with His147 in the HIE state. Structures were optimised with TPSS/def2-SV(P) and energies were obtained with TPSS-D3/def2-TZVPP. \label{cu-o_his_h-transfer_hie}}  
\end{figure}

We start by considering 
the \ce{C4-H} abstraction from the Cu--oxyl complex (\textbf{6b}$\rightarrow$\textbf{7a}). The structures of the reactant, transition state and  product as well as the QM/MM activation and reaction energies 
 are shown in Figures \ref{cu-o_his_h-transfer_hid} (with His147 in the HID state) and \ref{cu-o_his_h-transfer_hie} (with His147 in the HIE state). The energies shown here were calculated with TPSS, while additional details and the corresponding results with B3LYP are given in the SI, Table S6 and Figure S10. 
Independent of the His147 state, the \ce{Cu-O} distance of the oxyl ligand is remarkably constant, with a difference of only 0.04 {\AA} between the reactant and the product. The Cu--O bond length to the  tyrosine ligand (\textit{trans} to the oxyl group) also shows only minimal changes, from 2.4 {\AA} in the reactant to 2.5 {\AA} in the product. For the HID form, the activation energy of 69 kJ/mol and the overall reaction energy of $-17$ kJ/mol (73 and $-22$ kJ/mol with the B3LYP functional) show that the reaction is feasible. Interestingly, the HIE form gives rise to a higher activation energy, 104 kJ/mol, while the reaction energy is $-22$ kJ/mol (111 and $-27$ kJ/mol with B3LYP).
\begin{figure}[tbh!]
\centering
  \includegraphics[scale=0.28]{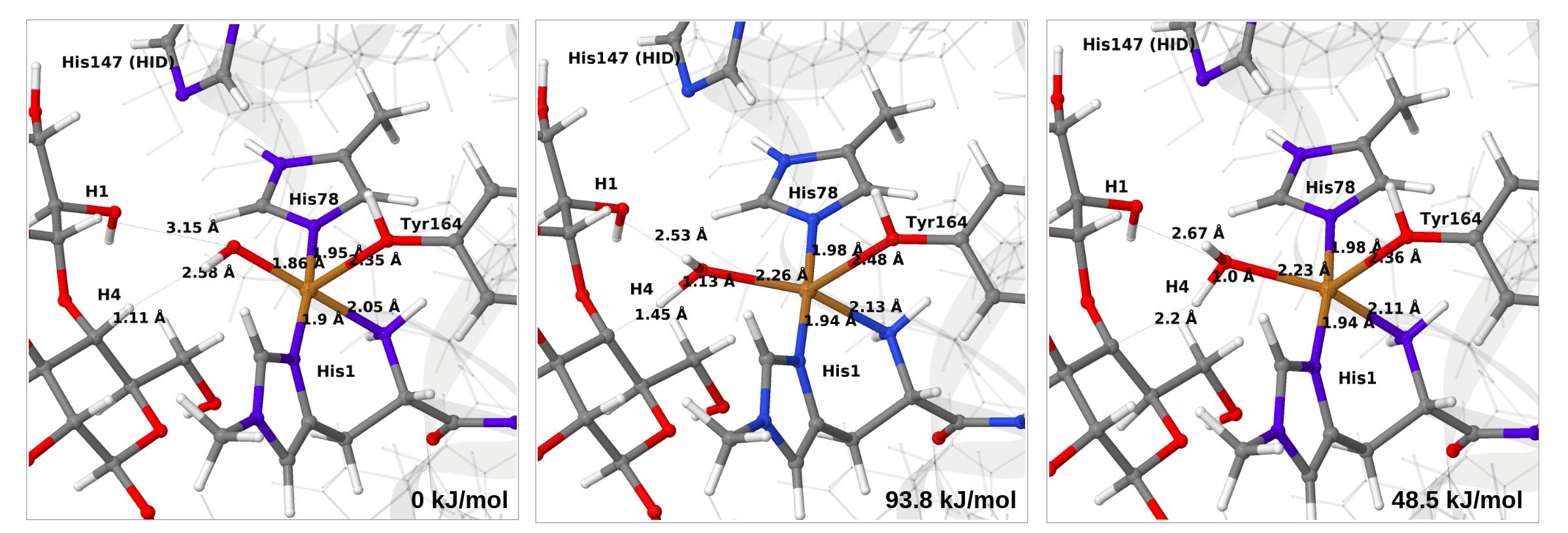}  
 \caption{Reactant \textbf{6c} (\textit{left}), transition state (\textit{middle}) and product \textbf{7b} (\textit{right}) for the \textbf{6c}$\rightarrow$\textbf{7b} reaction with His147 in the HID state. Structures were optimised with TPSS/def2-SV(P) and energies were obtained with TPSS-D3/def2-TZVPP. \label{cu-oh_his_h-transfer_hid}}  
\end{figure}
\begin{figure}[tbh!]
\centering
  \includegraphics[scale=0.28]{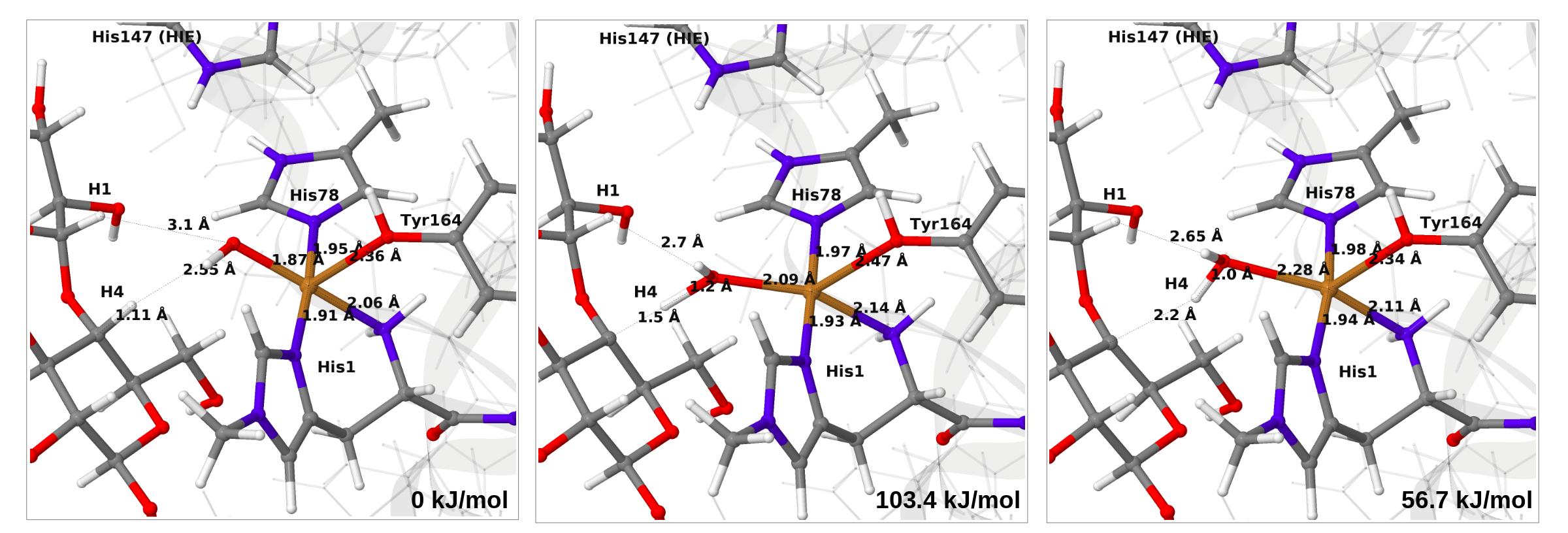}  
 \caption{Reactant \textbf{6c} (\textit{left}), transition state (\textit{middle}) and product \textbf{7b} (\textit{right}) for the \textbf{6c}$\rightarrow$\textbf{7b} reaction with His147 in the HIE state. Structures were optimised with TPSS/def2-SV(P) and energies were obtained with TPSS-D3/def2-TZVPP. \label{cu-oh_his_h-transfer_hie}}  
\end{figure}

We next turn to the Cu--hydroxyl complex (\textbf{6c}$\rightarrow$\textbf{7b}). The reactant, transition state and product structures and corresponding energies are shown in Figures \ref{cu-oh_his_h-transfer_hid} and \ref{cu-oh_his_h-transfer_hie} (additional details are given in Figure S11 and Table S7). In contrast to the \textbf{6b}$\rightarrow$\textbf{7a} reaction, we obtain a large change in the \ce{Cu-O} distance, from 1.9 {\AA} to 2.2--2.3 {\AA}, reflecting a change from hydroxyl to water. 

The ground state of \textbf{6c} was found to be a singlet, but as is evident from Figure S11, we find that the singlet--triplet energy splitting gradually decreases during the \ce{C-H} abstraction (as expected): For the TPSS functional, the triplet becomes energetically  lower than the singlet for a \ce{C4-H} distance of 1.45 {\AA} (1.40 {\AA} with B3LYP) and for the product, the triplet is significantly more stable. In the following, we estimate the activation energy from the energy at the singlet--triplet crossing. 

Our calculations also show that the activation energy is less sensitive to the protonation state of His147 for the Cu--hydroxyl intermediate \textbf{6c} than for Cu--oxyl (\textbf{6b}). For His147 in the HID form, we obtain  a \ce{C4-H} activation energy of 93 kJ/mol with the TPSS functional (97 kJ/mol with B3LYP). The corresponding values for the HIE variant are 103 and 93 kJ/mol, respectively. These values are close to the activation energy obtained for \textbf{6b} with His147 is in the HIE form (104--111 kJ/mol).
 
  Interestingly, the reaction energies for the \textbf{6c}$\rightarrow$\textbf{7b} reaction show some functional dependence, which was not seen for the  \textbf{6b}$\rightarrow$\textbf{7a} reaction. For the HID form, the TPSS functional predicts the \textbf{6c}$\rightarrow$\textbf{7b} reaction to be uphill by 49 kJ/mol, whereas B3LYP predicts that the reaction is slightly downhill with $-22$ kJ/mol. Similar values are obtained for the state with His147 in the HIE147 form: 57 (TPSS) and $-14$ kJ/mol (B3LYP), cf.~Table S7.
  The reason for the increased functional dependence in reaction \textbf{6c}$\rightarrow$\textbf{7b} is that the most stable spin-state changes in the course of reaction. This can lead to uncertainties with DFT methods (see e.g.~Ref.\citenum{pierloot2011}). We currently investigate this reaction with more accurate methods based on multireference perturbation theory.  
Yet, within the accuracy provided by DFT-based QM/MM, we can conclude that in the HIE protonation state of His147, the Cu--oxyl and Cu--hydroxyl have comparable activation energies, whereas the lowest activation energy is obtained for the Cu--oxyl complex with His147 in the HID protonation state. With the current accuracy of the employed exchange--correlation functional, we can not exclude that the reaction energies of \textbf{6b} and \textbf{6c} are similar.  
  We note also that the electrostatic environment contributions are uniformly small (2--10 kJ/mol, see Figures S10 and S11), which is in marked contrast to the much more sensitive proton-transfer reactions discussed above. 
 
To investigate the relative stability of \textbf{6b} and \textbf{6c}, we finally studied the protonation of \textbf{6b} to \textbf{6c}. Here, His147 acted as the proton donor and hence was in the HIP form. Protonation of \textbf{6b} turned out to be facile with a low energy barrier (below 20 kJ/mol) and with a overall reaction energy of around $-150$ kJ/mol. Therefore, \textbf{6b} should be rapidly converted to \textbf{6c} in acidic media.  Considering the fact that \textbf{6c} is very easy to generate, this intermediate should be considered the favored one, despite its slightly higher activation energy. 

\subsection{Recombination step and product formation}

\begin{figure}[htb!] 
\centering  
\includegraphics[scale=0.27]{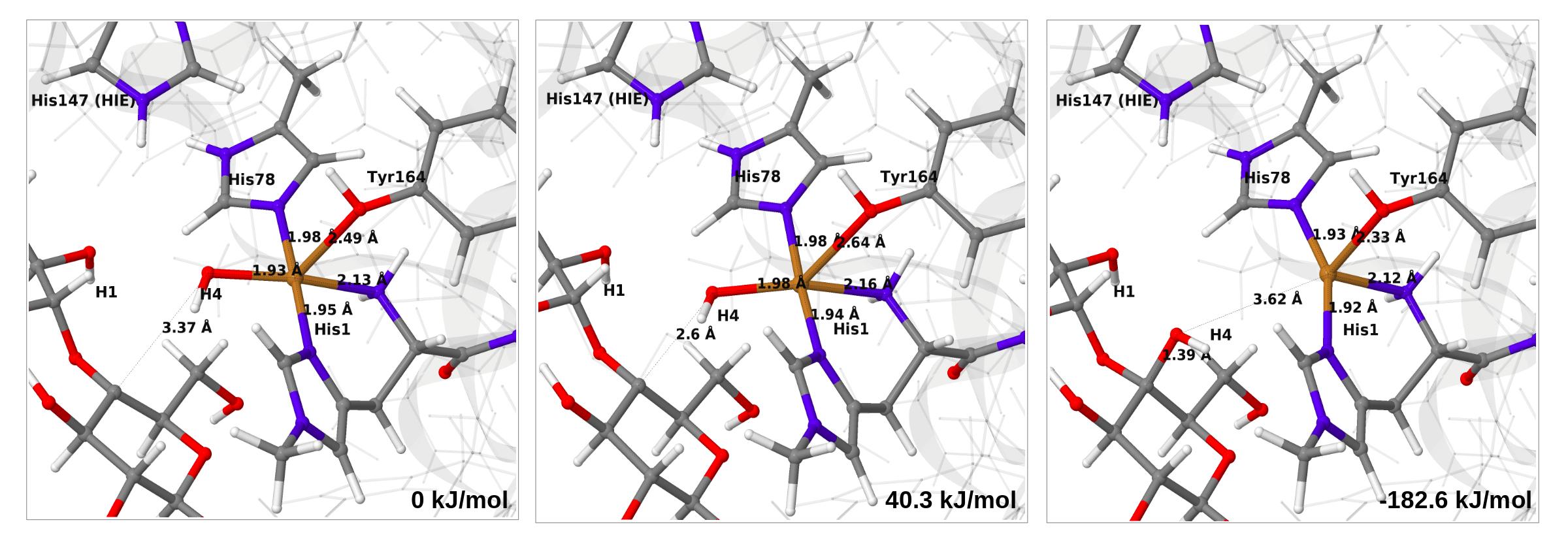} 
 \caption{Reactant \textbf{7a} (\textit{left}), transition state (\textit{middle}) and product \textbf{8a} (\textit{right}) for the  \textbf{7a}$\rightarrow$\textbf{8a} reaction with His147 is in the HIE state. The structures were optimised with TPSS/def2-SV(P) and energies were obtained with TPSS-D3/def2-TZVPP. \label{cu_oh_recomb_hie}  }  
\end{figure}

\begin{figure}[htb!] 
\centering  
\includegraphics[scale=0.27]{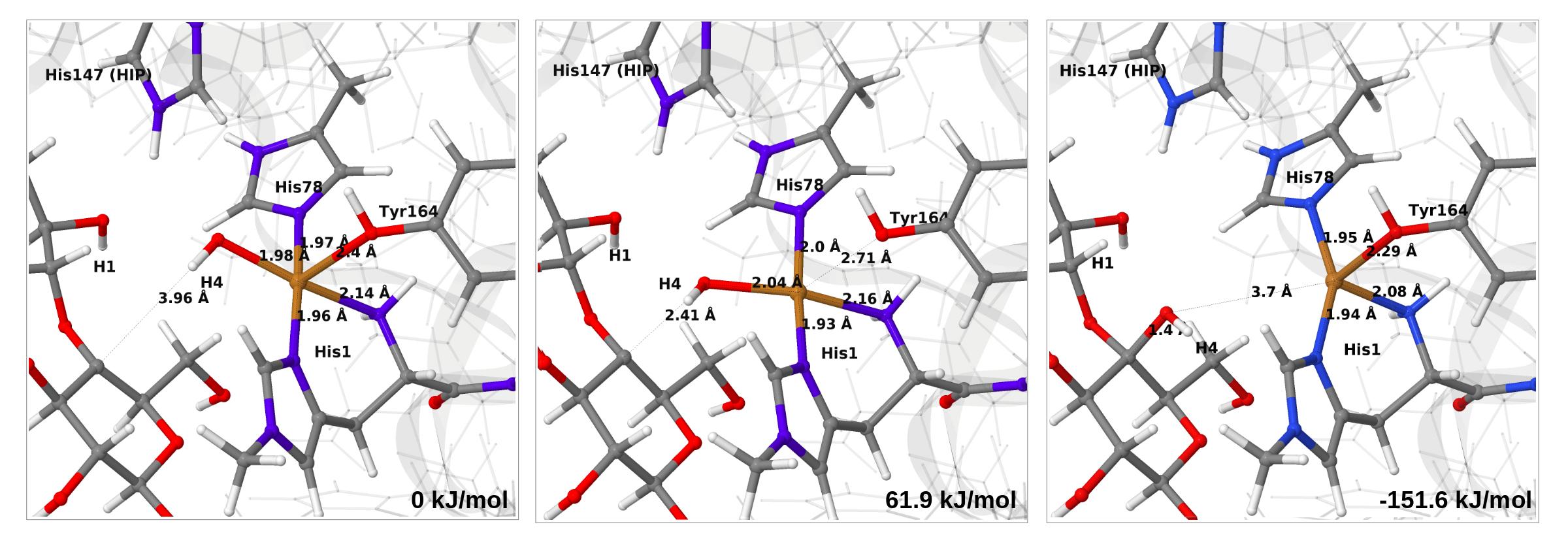} 
 \caption{Reactant \textbf{7a} (\textit{left}), transition state (\textit{middle}) and product \textbf{8a} (\textit{right}) for the \textbf{7a}$\rightarrow$\textbf{8a} reaction with His147 is in the HIP state. The structures were optimised with TPSS/def2-SV(P) and energies were obtained with TPSS-D3/def2-TZVPP.  \label{cu_oh_recomb_hip}  }  
\end{figure}

The final reaction is the recombination of the saccharide radical and the Cu ligand, i.e. the \textbf{7a}$\rightarrow$\textbf{8a} or \textbf{7b}$\rightarrow$\textbf{8b} reactions in Figure \ref{lpmo_active_site}. 
We first studied the reaction from \ce{[CuOH]^{+}} (\textbf{7a}). As noted in previous section, the HIE state is 121 kJ/mol more stable than the HID state, and we therefore focus on that state. However, as in previous section, we investigated  both tautomers and in this case the HIE and HID states give rise to similar activation and reaction energies. We also carried out the \textbf{7a}$\rightarrow$\textbf{8a} reaction with His147 in the HIP state. 
The reactant, \textbf{7a}, the transition state and product \textbf{8a} are shown in Figures \ref{cu_oh_recomb_hie} (for His147 in the HIE form) and \ref{cu_oh_recomb_hip} (for His147 in the HIP form) together with the  corresponding energies (see Tables S8--S9 and Figure S12 for further details). 
The reaction involves a transition from the triplet spin-state of \ce{[Cu-OH]+} and \ce{R$^{\bullet}$} (\textbf{7a}) to a closed-shell singlet state of the product (\ce{Cu+} and \ce{ROH}; \textbf{8a}). The transition state is the triplet--singlet crossing point. With His147 in the HIE form, the transition state has a \ce{C4-O} bond length of 2.6 {\AA} and the activation energy is 40 kJ/mol. With the B3LYP functional, the transition state occurs at a somewhat shorter bond distance (2.2 {\AA}) and the activation energy is slightly higher (53 kJ/mol, see Table S8). Independent of the employed functional, the reaction is strongly downhill by 182--193 kJ/mol. 
Both activation and reaction energies are similar for the HID state (albeit generally 10 kJ/mol lower, cf.~Figure \ref{LPMO_reaction_final_values}). 
The corresponding activation energy for the HIP state is either slightly higher (62 kJ/mol with TPSS) or similar to that with the HIE form (B3LYP), while the reaction is still significantly downhill (158--165 kJ/mol), independent of the employed functional.   

As expected, attempts to obtain a protonated \ce{ROH2+} product state directly from \textbf{7b} were unsuccessful. They were carried out with His147 in the HID state to allow the possibility for the product to protonate this residue, but this did not occur. 
However, the reaction (\textbf{7b}$\rightarrow$\textbf{8a} under protonation of HID to HIP) is downhill by between 43 and 47 kJ/mol, depending on the functional.  
Therefore, it may occur with a concerted protonation of HID to HIP under the transfer of \ce{OH$^{\bullet}$}. 
Since the active site is close to the surface, the deprotonation may also occur through connections to the bulk solvent. Another possible reaction path for \textbf{7b} could be protonation of His147 in the HID form to form \textbf{7a} + HIP. We are currently investigating these different possibilities.   

\section{Discussion}

\begin{figure}[htb!] 
\centering  
\includegraphics[scale=0.70]{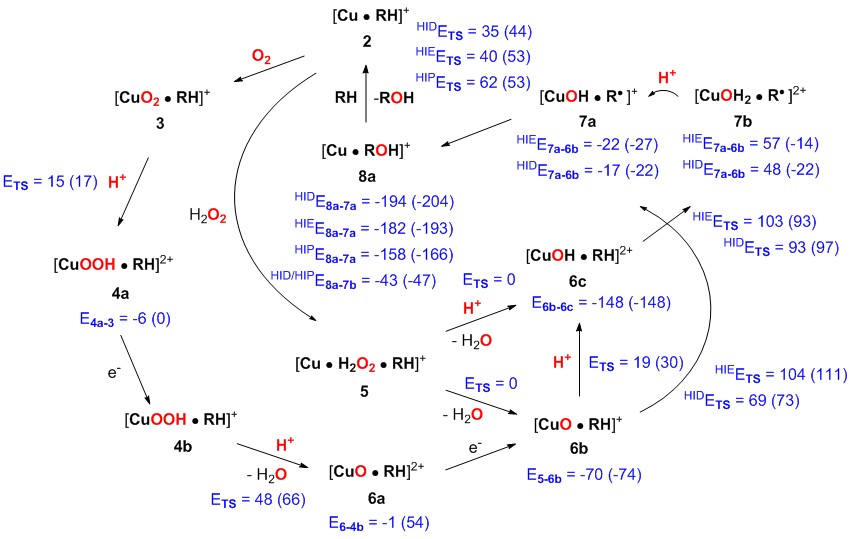} 
 \caption{Final, calculated mechanism for LPMO \ce{C-H} activation.   Activation energies are given as $E_{\text{TS}}$ and individual reaction energies are given below products, using the reactant as reference (all energies are in kJ/mol). The enegies are from TPSS with B3LYP in parentheses (always with def2-TZVPP basis sets). Protons are in all cases from His147 (in the HIP form). A few reactions are calculated with His147 in HIE/HIP or HIE/HID forms, as indicated with a subscript. No energies are reported for reactions involving a reduction of the reactant, and for \textbf{5}$\rightarrow$\textbf{6c}, because \textbf{6c} formed spontaneously.  \label{LPMO_reaction_final_values} }  
\end{figure}

We have calculated reaction and activation energies for the full putative mechanism of the LPMOs in Figure \ref{lpmo_active_site}. In this section, we  relate our findings to known experimental data and previous theoretical results. Our suggested mechanism is shown in Figure \ref{LPMO_reaction_final_values} along with the calculated reaction and activation energies. 

It is known that the investigated LPMO is C4-specific. Focusing on the structural parameters, we note that for our optimised structure of  the \ce{[CuO2]+} (\textbf{3}) complex,  the distances from \ce{O2-} to 
the H$_1$ and H$_4$ atoms of the substrate are 2.29~\AA~and 2.28~\AA, respectively (Figure \ref{cu_o2_qm_structures} and Table S1). These distances are slightly longer with the larger def2-TZVPD basis set (2.35 {\AA} and 2.41 {\AA}, Table S1).  The similarity of these distances suggests that a regiospecificity (for \ce{C4-H}) can only occur if the \ce{O-O} bond is broken before the hydrogen-transfer step, as tentatively suggested by O'Dell et al.\cite{dell2017}, based on superimposing a substrate-free structure of the \ce{[CuO2]+} complex with the structure of the LPMO--substrate complex\cite{frandsen2016}. 
 Interestingly, the difference between the two distances become larger with His147 in the HIP form, making the \ce{C4-H} abstraction less likely. 
 The distances between the \ce{[CuO]+} oxygen atom in \textbf{6b} and the \ce{H4} and \ce{H1} hydrogen atoms in the substrate are 2.4 {\AA} and 3.1 {\AA}, respectively. 
 Similarly, the distance between the \ce{[Cu-OH]^{2+}} oxygen atom and \ce{H4} and \ce{H1} hydrogen atoms are 3.6 {\AA} and 4.2 {\AA}, respectively for \textbf{6c} (singlet). Thus, the  difference in \textbf{6b} and \textbf{6c} are 0.6--0.7 {\AA} (a similar difference is observed for the triplet spin-state of \textbf{6c}), which is much larger than the 0.1--0.2 {\AA} for the corresponding \ce{[CuO2]+} intermediate (cf.~Figures \ref{cu_o2_qm_structures} and \ref{cu_o2_his147_h_transfer}).  Thus, we confirm that the structures after cleavage of the \ce{O-O} bond can explain the \ce{C4-H} regioselectivity of this LPMO. The calculated energies also support this; structures involving \ce{[Cu-O]+} (\textbf{6b}) and \ce{[Cu-OH]^{2+}} (\textbf{6c}) have activation energies of 93--111 kJ/mol (with His147 in the HIE form, Table S7), which are much lower than the 140--168 kJ/mol we obtain for \textbf{3} (Table S5; also with His147 in the HIE form). 
 
 Unfortunately, direct comparison of these activation barriers with experiment is not straight-forward, as kinetic data is difficult to obtain for LPMOs and has therefore been scarce. Rate constants $k_{\text{obs}}$\cite{agger2014,loose2014,walton2016,frandsen2016} between 0.01 s$^{-1}$ and 0.1 s$^{-1}$, and in some special cases\cite{cannella2016,loose2016} up to 0.2--0.5 s$^{-1}$ have been reported for different LPMOs. The LPMO we target in this study\cite{frandsen2016} has a $k_{\text{cat}}$ = 0.11 s$^{-1}$ (at 310 K), which (according to transition-state theory) translates into an activation free energy of 81 kJ/mol if the pre-exponential factor is set to $kT/h = 6\cdot 10^{12}$\cite{jensen2007}. Note also  that Ref.~\citenum{walton2016} translates the rate constant to 63 kJ/mol, presumably due to a different pre-factor. 
 
 We find that the \ce{C-H} abstraction is the rate-determining step. We  have in our previous study\cite{hedegaard2017b} (using a smaller model) obtained a thermochemical energy correction of $-6$ kJ/mol for \textbf{6b} and $-13$ kJ/mol for \textbf{6c}. Assuming a similar magnitude of correction for the activation barrier, we obtain activation energies of 63 and 81 kJ/mol for \textbf{6b} and \textbf{6c} and His147 in the HID state (these values are with the TPSS functional; the corresponding B3LYP values are 67 and 84 kJ/mol). The corresponding values with His147 in the HIE state are in most cases slightly higher: 99 kJ/mol for \textbf{6b} and 90 kJ/mol for \textbf{6c} with the TPSS functional (105 and 80 kJ/mol with B3LYP).  The approximation of using a thermodynamical correction from reaction energies for barrier heights deserves a comment. The correction mainly comes from the zero-point vibrational energy (ZPVE) and it is of the same magnitude as literature values on barrier heights for H-abstractions. For instance, the authors of Ref. \citenum{wang2018} estimated the effect to be $-$11 kJ/mol from ZPVE and 7 kJ/mol from entropy ($T\Delta S$), resulting in a total correction of $-$4 kJ/mol. However, it should be noted that these numbers are taken from an organic system. Thiel and co-workers\cite{senn2009b} showed for the H-abstraction in a metalloenzyme (P450cam) that ZPVE can be somewhat higher ($-$18 kJ/mol), whereas the corrections from entropy indeed are small (below 5 kJ/mol). Nevertheless, with an estimated magnitude of 5--18 kJ/mol, the choice of functional quite often introduces larger errors.  

With these considerations in mind, the calculated activation energies show that both \textbf{6b} and \textbf{6c} are likely intermediates and with His147 in the HID form, they give activation energies close to the experimental one. With His147 in the HIE form, the activation energies are in most cases slightly larger than the experimental one. In a recent study, the same LPMO as studied here was shown to be active with a number of different polysaccharide substrates, and it was suggested that different mechanisms were operative for different substrates.\cite{simmons2017} Our study shows that even for the same substrate, different active species can be operative, depending on the second coordination sphere.  

Our calculations thus show that His147 may have several roles in the mechanism. 
First,  it is a possible proton donor, which can explain the results from recent mutagenis experiments, showeing that mutation of this residue results in a decreased activity of LPMO\cite{span2017}.
 Second, this histidine residue is, together with Gln162, involved in positioning \ce{H2O2} for dissociation. Finally, 
the  protonation state (and tautomeric from) of His147 also influences the activation energy of the \ce{C-H} step, which may alter which species is most reactive. Notably, both His147 and Gln162 are highly conserved among different AA9 LPMOs.\cite{span2017}  

Given the large sequence variance between LPMOs (no residues, besides those involved in the histidine brace, are strictly conserved in all LPMO classes, AA9, AA10, AA11 and AA13), it is interesting to speculate in how general our suggested mechanism is, especially as that the second coordination sphere often plays a role. Since His147 and Gln162 are generally conserved in AA9 LPMOs, it is likely that the suggested mechanism is valid for AA9 LPMOs. The AA10 LPMOs vary considerably depending on their substrate specificity (chitin or cellulose).\cite{forsberg2014b} Chitin-active AA10 LPMOs have a Glu residue in place of Gln162, and it is possible that this residue might play the combined role of His147 and Gln162 (i.e. proton donor and positioning of \ce{H2O2}). Similarly, both AA11\cite{hemsworth2014} and AA13\cite{leggio2015} have a Glu/Asp and Gln in the in second coordination sphere, but the number of characterized structures for these classes are lower. Cellulose-active AA10 are again different, but has concerved  H-bonding mortifs that resemble the His147 and Gln162 mortif.\cite{forsberg2014a,beeson2015} For instance, in the cellulose-active AA10 LPMO with the PDB entry 4OY7,\cite{forsberg2014a} a Glu residue replaces Gln162 and an Arg residue replaces His147.\cite{forsberg2014a} Yet, it is clear that further investigations are required to fully understand the role of the second coordination sphere in the mechanism.

While the good agreement between the calculated and experimental activation energies are reassuring, activation energies obtained from experimental rate constants should also be interpreted with some care. For instance, the LPMO rate constant have shown to depend on the externally applied reductant\cite{loose2016}, as discussed further below. 

It is notable that a previous QM-cluster study\cite{kim2014} also suggested an oxyl species, but this study is not directly comparable to ours, because \ce{O2-} and \ce{O$^{\cdot}$-} were positioned \textit{trans} to the tyrosine ligand, based on initial (and probably incomplete) crystallographic data\cite{li2012}. Both our previous QM/MM study\cite{hedegaard2017} and a recent experimental study\cite{dell2017} suggest that \ce{O2} binds in an equatorial position.
In Ref.~\citenum{kim2014}, they obtained activation barriers of 146 kJ/mol and 64 kJ/mol for species with \ce{O2-} and \ce{O$^{\cdot}$-}, respectively (using the B3LYP functional). Accordingly, our results agree with Ref.~\citenum{kim2014} that the oxyl-species is more reactive than the superoxide. Very recently, a QM-cluster study\cite{bertini2017} and a QM/MM study\cite{wang2018} (employing a equatorial coordination of oxyl and superoxo ligands) have also suggested the oxyl species to abstract the \ce{C-H} bond, although Ref.~\citenum{wang2018} finds that hydroxylation of the glycoside bond has a higher activation energy. 

Most evidence thus seems to point towards that hydrogen abstraction from \ce{C-H} occurs after \ce{O-O} bond breaking, either with a \ce{[Cu-O]+} or a \ce{[Cu-OH]^{2+}} active species or both.  With the present results, it is naturally interesting to compare the two investigated mechanisms for generation such species. Starting with the \ce{O2} pathway in Figure \ref{lpmo_active_site}, the initial protonation of the superoxide \textbf{3}$\rightarrow$\textbf{4a} is certainly feasible with an activation energy below 20 kJ/mol. 
 The cleavage of the O--O bond in \textbf{4a} requires further reduction to \textbf{4b} and our calculations suggest that the second reduction (of  \textbf{4b} in Figure \ref{lpmo_active_site}) does not occur before the bond is cleaved. Instead, the O--O bond is cleaved in \textbf{4b} to generate an oxyl species (\textbf{6a}) with an activation energy of 48 (TPSS) to 66 kJ/mol (B3LYP), which is slightly lower than for the \ce{C-H} abstraction step. The reaction is thermoneutral with TPSS but uphill with B3LYP, showing that \textbf{6a} is expected to have a short life time. Thus, these calculations show that the oxyl and hydroxyl species \textit{can} be generated from \ce{O2}, but it requires an external supply of electrons and their timely delivery to the \textbf{4a} and \textbf{6a} intermediates. The short life-time of \textbf{6a} makes it a unlikely candidate for the reactive species for the \ce{C-H} abstraction, although the bond-dissociation energy was favorable.\cite{hedegaard2017b}  
  
  The question on how these electrons are supplied has been much discussed. A number of different electron-transport systems have been demonstrated for LPMOs, ranging from the protein cellobiose dehydrogenase (CDH)\cite{phillips2011} to small redox-active molecules\cite{kracher2016} and even light-activated systems\cite{bissaro2016,cannella2016}. Still, the mechanism of the electron shuttle from donor to LPMO is unclear. For the initial reduction of the resting state (\textbf{1}), it has been shown that CDH can transfer electrons with rate constants that are larger than the LPMO reaction ($k_{obs} = $ 0.9--67 s$^{-1}$)\cite{tan2015,kracher2016}, which is difficult to combine with the observed dependence of the LPMO activity on the reductant.\cite{kracher2016,bissaro2017} A possible explanation could be that the reduction potentials of the transient species, e.g., \textbf{4a} and \textbf{6a}, are significantly different from that of \textbf{1}. Since calculations of reduction potentials are associated with large errors (as we have previously shown for LPMOs\cite{hedegaard2017b}),  we can at present not investigate this option in detail. 
Currently it is known that the initial reduction (\textbf{1}$\rightarrow$\textbf{2}) does not need to be carried out in presence of substrate\cite{tan2015}, but it is more difficult to envisage how CDH (or other reductants) transfers the electrons when the substrate shields the Cu active site during reaction, although long-range electron transfer\cite{westereng2015} and electron-transfer chains through the LPMO protein\cite{li2012} have been suggested.  
Since our calculations show that generation of active Cu-oxygen species is possible via the \ce{O2} pathway in Figure \ref{lpmo_active_site}, this pathway must still be considered a viable option, although it is not our favored pathway.

The \ce{H2O2} pathway alleviates the need for transfer of electrons when the substrate is bound to the active site.\cite{bissaro2017} It is known that LPMOs can produce \ce{H2O2} in absence of polysaccharide susbtrate\cite{kittl2012,bissaro2016}, although the exact concentration is difficult to estimate.\cite{bissaro2017} Our calculations support the \ce{H2O2} pathway, as both \textbf{6b} and \textbf{6c} are readily generated from \ce{H2O2}, probably through a state with \ce{H2O2} bound in the second coordination sphere of Cu (\textbf{5}), because no states with \ce{H2O2} coordinated to Cu(I) could be located. Note that binding of \ce{HO2-} to Cu(II) is also a possibility, forming  \textbf{4b}, meaning that the "priming" reduction could be reduction of \textbf{4b}, combined with cleavage of the O--O bond, giving rise to one of the \textbf{6} species (which could explain the observed dependency on the reductant). While this scenario might be worth further investigation, it must currently be considered speculative. 

Neither  Ref.~\citenum{kim2014} nor Ref.~\citenum{bertini2017} considered the hydroxyl intermediate or how the reactive species are generated, although Ref.~\citenum{bertini2017} did obtain a structure in which \ce{H2O2} coordinates to Cu(I). However, the  structure was optimised without the substrate, using a different crystal structure. 
The generation of an oxyl species from \ce{H2O2} (but not \ce{O2}) was considered in Ref.~\citenum{wang2018}: our prediction that formation of an oxyl complex is feasible from \ce{H2O2} generally agrees with the results from this study  (although Ref.~\citenum{wang2018} did not consider hydroxyl complexes). 

Finally, we note that it has recently been suggested that deprotonation of the terminal \ce{NH2} unit may be part of the mechanism\cite{walton2016,dhar2015}, but  no  conclusive evidence for this deprotonation has yet been published. A recent neutron and X-ray diffraction study has been claimed show such a state,\cite{bacik2017} but the crystallographic data are ambiguous: In a separate study we have re-investigated this crystal structure using quantum refinement. We see no evidence of a terminal \ce{NH2} deprotonation and this path is therefore not considered here. Moreover, our QM-cluster studies showed that deprotonation did not lead to any enhanced reactivity\cite{hedegaard2017b}.

\section{Conclusions}

In this paper, we have investigated the substrate--LPMO complex with QM/MM methods, starting from a recently published crystal structure\cite{frandsen2016}. The full mechanism has been mapped out and the results are shown (together with activation and reaction energies) in Figure \ref{LPMO_reaction_final_values}. 

 The calculations show that protonation of a superoxo \ce{[CuO2]+} complex is feasible and can lead to formation of an oxyl complex after reduction and dissociation of water. This oxyl complex is readily protonated to a Cu--hydroxyl complex. Both Cu--oxyl and Cu--hydroxyl (\textbf{6b} and \textbf{6c}) are sufficiently reactive to abstract a hydrogen from the polysaccharide substrate. 
 In addition, we have shown that the protonation state of the second-sphere histidine (His147) can shift the balance between two intermediates by enhancing the reactivity of  the Cu--oxyl intermediate. 
 
In a parallel set of calculations we have investigated the generation of the two reactive intermediates (\textbf{6b} and \textbf{6c}) from  \ce{H2O2} and our calculations show that this is more favorable. Thus, the calculations support \ce{H2O2} as co-substrate.  

All in all, the current calculations shows a new route to LPMO activity, through both Cu-oxyl and Cu--hydroxyl intermediates. They further support \ce{H2O2} as co-substrate and pin-point the role of the second coordination sphere.
In future calculations, we intend to address the \ce{C-H} abstraction with more accurate multireference methods and to address the problem of calculating reduction potentials for the various states of LPMO. It will furthermore be interesting to compare the calculated rate for generation of \ce{H2O2} without substrate\cite{kjaergaard2014}, which can be directly compared to the rate-determining \ce{C-H} abstraction step.  By such a calculation, the relative importance of the generation of \ce{H2O2} from LPMOs can be compared to the enzymatic reaction. Finally, a recent paper showing that particulate methane monooxygenase has a mono-nuclear active site\cite{cao2017} indicates  that our conclusions may be generalized also to that enzyme. We are currently investigating this possibility.


\begin{acknowledgement}

This investigation has been supported by grants from the Swedish research council (project 2014-5540) and from COST through Action CM1305 (ECOSTBio). EDH thanks the Carlsberg foundation (CF15-0208) and the European Commission for support. The computations were performed on computer resources provided by the Swedish National Infrastructure for Computing (SNIC) at Lunarc at Lund University.

\end{acknowledgement}


\bibliography{lpmo}

\end{document}










\section*{Computational procedures and setup}

\subsection*{Protein setup \label{setup}}

The starting coordinates where taken from the 1.8 {\AA} resolution crystal structure of the LPMO--substrate complex from \textit{Lentinus similis}\cite{frandsen2016} (PDB ID  5ACF). The structure is a monomer that contains 250 amino acids and 358 crystal water molecules, amounting to 2216 atoms in total. The structure was collected with a low radiation dose to minimise photoreduction during data collection, which is a well-known problem for metalloproteins\cite{sommerhalter2005} like LPMOs\cite{hemsworth2013b,gudmundsson2014}.  

The crystal structure contained one amino-acid residue (Glu235) and ten water molecules with alternative conformations 
We included in the calculations only the conformation with highest occupation or the first conformation if the occupation numbers were equal. 
It also contained ten \ce{Cl-} ions (one coordinated to Cu) and an N-acetylglucosamin (NAG) group bound to Asn138, which were deleted.   
Hydrogen atoms were added using the Maestro protein preparation tools.\cite{Maestro2015} For titratable residues (7 arginine, 3 lysine, 7 histidine, 16 aspartate and 5 glutamate residues), Maestro employs the PROPKA program\cite{olsson2011} to estimate pK$_{\text{a}}$ values. 
The individual residues were visually inspected and their solvent exposure and hydrogen-bond network were assessed.
Based on this, we concluded that all arginines and lysines are protonated (+1) and the aspartic and glutamine acids are in their carboxylate form ($-1$). 
Three of these charged residues are buried inside the protein, Glu103, Arg140 and Glu142. Glu103 forms an ionic pair with the ammonium group of Lys100, whereas Arg140 and Glu140 from another ionic pair. Therefore, they were considered in the charged forms.
The protein contains two cysteine residues (Cys41 and Cys167) that are cross-linked by a disulfide bridge. 

Histidine residues have two possible protonation sites and 
in the following, we denote histidines as HIE (N$^{\epsilon 2}$ protonated), 
HID (N$^{\delta 1}$ protonated) or HIP (both nitrogens protonated).   
The N-terminal histidine is a special case, because the imidazole is 
methylated on the N$^{\epsilon 2}$ atom, whereas  N$^{\delta 1}$ coordinates to the Cu ion.
For the remaining histidine residues, we employed the protonation states 
HIP66, HID78, HIP79, HIP122, HID125, HIP131 and HIE147.  HIP66 
forms hydrogen bonds from N$^{\delta 1}$ to the O$^{\delta 1}$ carboxyl group of Asp72
and from N$^{\epsilon 2}$ to a hydroxy group of the substrate. 
HID78 coordinates to Cu through N$^{\epsilon 2}$, whereas HIP79 forms a salt-bridge from H$^{\delta 1}$ to the carboxylate group of Asp116 and a hydrogen bond to a crystal water molecule through H$^{\epsilon 2}$. 
HIP122 forms a hydrogen bond to the carboxyl group of Asp70 through H$^{\delta 1}$ and is exposed on the surface of the protein and was therefore chosen to be doubly protonated. 
HID125 forms a hydrogen bond to a crystal water molecule from H$^{\delta 1}$ and N$^{\epsilon 2}$ accepts a hydrogen bond from the NH group of the Trp64 side chain. 
HIP131 is solvent exposed on the surface and was chosen to be doubly protonated (in the crystal structure one of the hydrogen interacts with a \ce{Cl-} ion). 
HIE147 is close to the active site and may participate in the reaction mechanism as the proton donor. The preperation tool in Maestro originally flipped this residue, but we decided instead to employ the HIE form in the conformation obtained in the crystal structure.  This form is in agreement with a recent neutron diffraction study\cite{dell2017} on a substrate-free structure, which allows H$^{\epsilon 2}$ to interact with the \ce{O2-} ligand in \ce{[CuO2]+}.  
We used this state during setup and equilibration, whereas in the QM/MM calculations we employed sometimes instead the other two states to model various proton-transfer events. 
With His147 in the HIE state, the total charge of the simulated system in the \ce{[Cu(H2O)]^{2+}} (\textbf{1}) resting state was -5.

\subsection*{Equilibration and QM/MM setup}  
The system described above was equilibrated by simulated annealing. 
Both the equilibration and the QM/MM calculations followed closely our previous investigations\cite{hedegaard2017} and here we only highlight differences.
The previous calculations were carried out on a system without substrate, but in this study, we included a trisaccharide substrate (cf.~Figure 1 in the paper). 
The substrate was described by the glycam.v06 force field,\cite{kirschner2008} which is tailored for oligosaccharides. 
The protein was described with the Amber FF14SB force field \cite{maier2015} and water molecules with the TIP3P model.\cite{Jorgensen1983}
The equilibration was performed on state \textbf{1}, which was obtained by replacing the coordinating \ce{Cl-} ion in the crystal structure with \ce{H2O}.
\begin{figure}[tbh!]
\centering
  \includegraphics[scale=0.40]{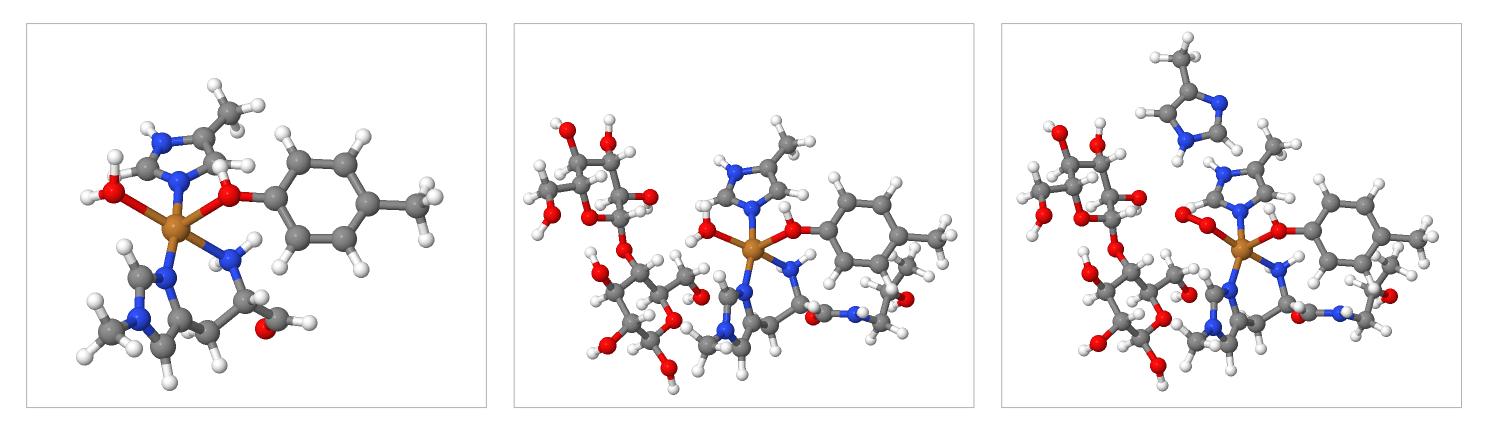}
 \caption{The systems employed for RESP charges (\textit{left}) and two examples of the employed QM systems (\textit{middle} with and \textit{right} without His147; intermediates \textbf{1} and \textbf{3} are used as examples).     \label{qm_systems} }  
\end{figure}

For the equilibration, restrained electrostatic potential (RESP) charges were employed for the metal center and its first coordination sphere (cf. Figure \ref{qm_systems})
The employed structure was taken from the crystal structure 
and only the hydrogen atoms were optimized, employing the TPSS functional\cite{tao2003} together with 
the def2-SV(P) basis set\cite{bs920815sha,eichkorn1997}. 
The electrostatic potential was calculated in points that were sampled with the Merz--Kollman scheme\cite{singh1984,besler1990} using default radii for the light atoms and 2~{\AA} for Cu\cite{sigfridsson1998}.
They were employed by the RESP program (a part of the AMBER software package) to calculate the charges.

The QM system (system 1) consisted of the copper ion and its first coordination sphere i.e. the imidazole ring of His78 and the phenol ring of Tyr164, both 
capped with a hydrogen atom replacing C$^{\alpha}$. The entire His1 residue, which coordinates to 
Cu through the terminal amino group as well as the imidazole side chain, was included. The neighboring Thr2 residue was included up to the C$^{\alpha}$ atom, which was replaced by a hydrogen atom.  
In addition, the two first glucose rings of the substrate were also included in system 1, whereas the third glucose unit was described by MM. 
Thus, for the \textbf{1} state, the QM region comprised 111 atoms (see Figure \ref{qm_systems}, \textit{middle}).  

Since the reduction of \textbf{1} without substrate has been 
discussed in several previous computational QM-cluster\cite{kim2014,kjaergaard2014,gudmundsson2014} and QM/MM studies\cite{hedegaard2017},  we have also considered the reduced \ce{[Cu(H2O)]+} state (\textbf{2}). From the optimised structure of \textbf{1} we constructed the \ce{[CuO2]^{+}} (\textbf{3}) state by replacing  
the equatorial \ce{H2O} with \ce{O2^{$\bullet$}-}. Starting from \textbf{3}, we also included His147 within the QM region (Figure \ref{qm_systems}, \textit{right}) in various states of protonation (HID, HIE or HIP), depending on the intermediate. For reactions where His147 acts as proton donor, we employed the HIP state.  For the active intermediates in \ce{C-H} activation, we studied both HIE and HID forms. 

The QM/MM structure optimizations employed the dispersion-corrected TPSS-D3 functional\cite{tao2003,grimme2010} with Becke--Johnson damping\cite{grimme2011} and the def2-SV(P) basis set\cite{bs920815sha,eichkorn1997}.  
The \ce{[CuO2]+} and \ce{[CuO]+} intermediates have low-lying singlet states, which are best calculated as an open-shell singlet with the broken-symmetry approach.\cite{noodlemann1986} All species with an even number of electrons have been calculated both as triplet and singlet species. The most stable state was employed, but small singlet--triplet splittings are commented. All reported energies were obtained from TPSS-D3 single-point calculations (with the full protein represented by point-charges) with the def2-TZVPP basis set\cite{bs920815sha} on structures obtained with TPSS-D3/def2-SV(P).  In addition, we also report energies for similar calculations with TPSS-D3 replaced with B3LYP-D3.\cite{becke1988,becke1993,lee1988}

For selected states (\textbf{1}--\textbf{3}), we have probed the quality of the structures obtained with the TPSS-D3 functional and def2-SV(P) basis set by increasing the basis set to def2-TZVPD. In addition, test calculations were also performed with system 2 optimised. System 2 was defined as all atoms within 6 {\AA} of any atom in system 1. We use the labels ``free'' and ``fixed'' for calculations in which the coordinates of atoms in system 2 were or were not optimized, respectively. 

\section*{Overlay of active sites in 5ACF and 4EIS crystal structures }

\begin{figure}[htb!]
\centering
  \includegraphics[scale=0.42]{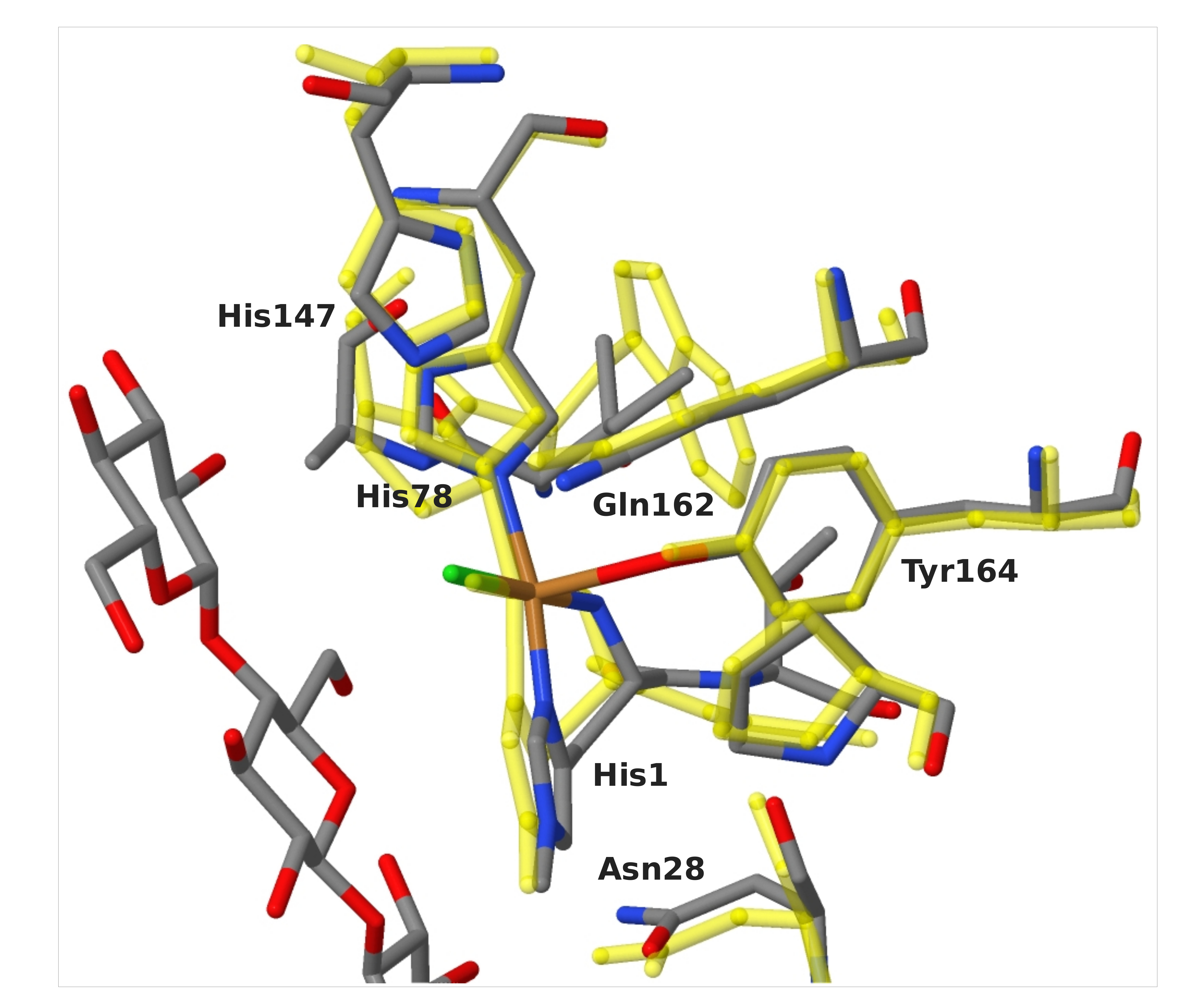}
 \caption{ Overlay of active site structures from 5CAF and 4EIS (yellow). Hydrogen atoms are not shown. Residues numbering is accoriding to 5CAF. \label{overlay_with_resnum} }  
\end{figure}
\newpage

\section*{Optimisations with system 2 free}

\begin{table*}[htb!]
\centering
\caption{Cu--ligand bond lengths (\AA) in the present and in previous studies for \textbf{3} with substrate (entries 1--3 and 9--11) and without substate (entries 4--8 and 12,13). All QM/MM results were obtained with the TPSS-D3 functional and def2-SV(P) basis set (unless otherwise noted). \label{opt_CuOO_state_eq} }
{\tiny
\begin{tabular}{lllccccccc}
 \hline\hline \\[-2.0ex]
Entry &Spin  & Source  & System 2  &  \ce{Cu-N$^{\epsilon}$_{His78}} &  \ce{Cu-N}$_{\text{His1}}$  & \ce{Cu-N}$^{\epsilon}_{\text{His1}}$ &  \ce{Cu-O}$_{\text{Tyr}}$ & \ce{Cu-O}$_{2}$ &   \ce{O-H1}/\ce{O-H4}    \\[0.5ex]
\hline \\[-1.5ex]  
1 & 1        & QM/MM                      & fixed &  1.98  &     2.11     & 1.95   &  2.28     &  2.09            &  2.27/2.26  \\[0.5ex]
2 & 1        & QM/MM                      & free  &  2.01  &     2.12     & 1.98   &  2.31     &  2.09            &  2.29/2.28  \\[0.5ex]
3 & 1$^{a}$  & QM/MM                      & free  &  2.00  &     2.12     & 1.97   &  2.39     &  2.07            &  2.35/2.41  \\[0.5ex]
4 & 1        & QM/MM\cite{hedegaard2017}  & fixed &  2.06  &     2.15     & 2.01   &  2.89     &  2.04            &    -        \\[0.5ex]
5 & 1        & QM/MM\cite{hedegaard2017}  & free  &  2.06  &     2.13     & 2.00   &  2.84     &  2.04            &    -        \\[0.5ex]    
6 & 1$^{a}$  & QM/MM\cite{hedegaard2017}  & free  &  2.06  &     2.12     & 2.00   &  2.94     &  2.01            &    -        \\[0.5ex]
7 & 1$^{b}$  & QM/MM\cite{hedegaard2017}  & free  &  2.08  &     2.11     & 2.01   &  2.84     &  1.99            &    -        \\[0.5ex] 
8 & 1        & QM-custer\cite{kjaergaard2014}     & - &  1.98  & 2.09     & 1.97   &  3.35     &  1.98            &   -         \\[0.5ex]                                    
9 & 1        & QM-custer\cite{bertini2017}        & - &  1.97   & 2.16     & 1.95   &  4.14     &  2.11            &  3.02/2.38  \\[0.5ex]
\hline
\\[-2.0ex]
10 & 0       & QM/MM                       & fixed &  1.99  &    2.11      & 1.96   & 2.28      &  2.06            &  2.24/2.25 \\[0.5ex]
11 & 0       & QM/MM                       & free  &  2.01  &    2.11      & 1.99   & 2.31      &  2.06            & 2.26/2.27  \\[0.5ex]
12 & 0       & QM/MM\cite{hedegaard2017}   & fixed &  2.06  &    2.15      & 2.01   & 2.87      &  2.02            & -          \\[0.5ex]
13 & 0       & QM/MM\cite{hedegaard2017}   & free  &  2.06  &    2.12      & 2.00   & 2.82      &  2.03            & -          \\[0.5ex]  
\hline \hline  
\end{tabular} \\
$^a$ Optimised with TPSS-D3/def2-TZVPD $^b$ Optimised with B3LYP-D3/def2-TZVPD.
}
\end{table*}

\begin{figure}[htb!]
\centering
  \includegraphics[scale=0.42]{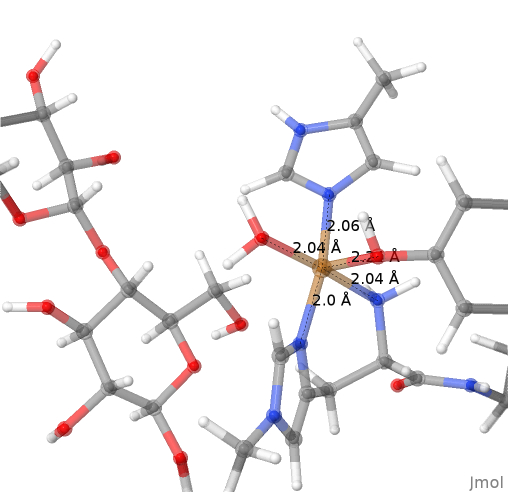}
    \hspace{1.0cm}
  \includegraphics[scale=0.42]{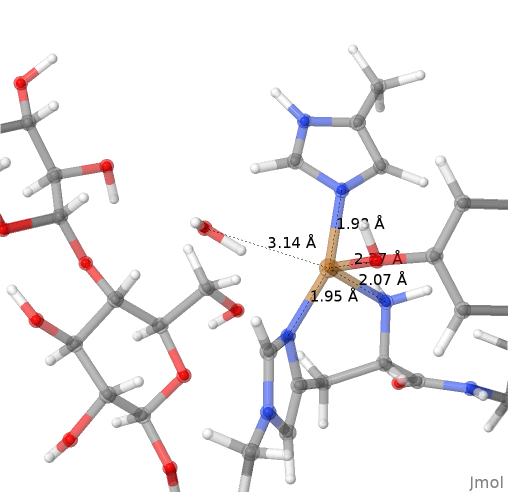}
 \caption{Optimised structures of \ce{[Cu(H2O)]^{2+}} (\textbf{1}, \textit {left}) and  \ce{[Cu(H2O)]+} (\textbf{2}, \textit{right}). The optimisation was carried out with TPSS-D3/def2-SV(P) with system 2 free.\label{cu-rest_state_red_fix_vs_free_2}  }  
\end{figure}
\begin{figure}[tbh!]
\centering
  \includegraphics[scale=0.48]{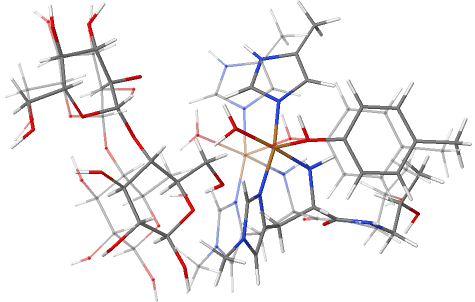}
\hspace{0.1cm}  
    \includegraphics[scale=0.48]{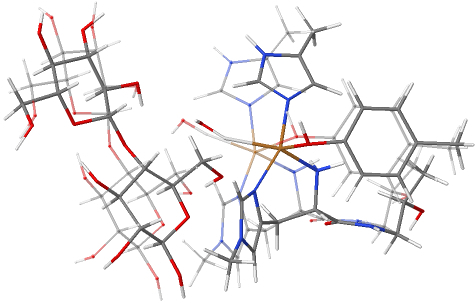}
 \caption{Comparison of optimised structures of \ce{[Cu(H2O)]^{2+}} (\textbf{1}, \textit{left}) and  \ce{[Cu(H2O)]^{+}} (\textbf{2}, \textit{right}) with system 2 fixed (thin lines) or free (thick lines). The optimisations were carried out with TPSS-D3/def2-SV(P).  \label{cu-rest_state_red_fix_vs_free} }  
\end{figure}
\begin{figure}[htb!]
\centering
  \includegraphics[scale=0.42]{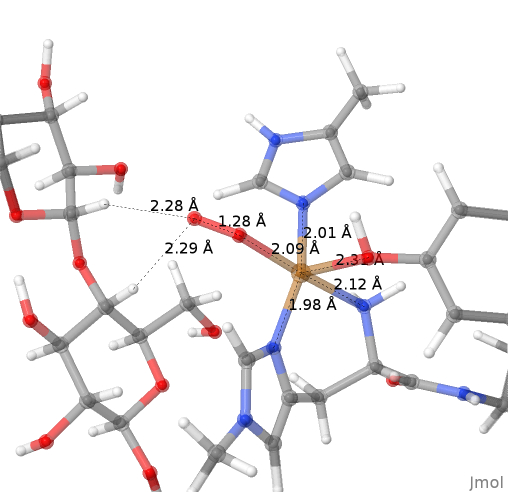}
      \hspace{1.0cm}
  \includegraphics[scale=0.42]{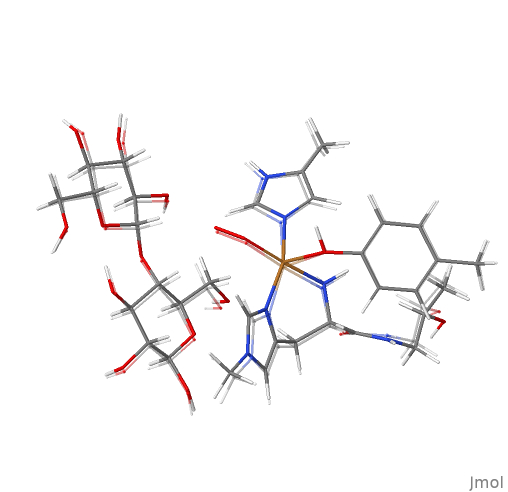}
 \caption{Optimised structure of \textbf{3}, obtained with TPSS-D3/def2-SV(P) and system 2 fixed (\textit{left}). The figure to the \textit{right} shows an overlay of \textbf{3} optimised with system 2 fixed (thick lines) and  free (thin lines).\label{cu_o2_qm_structures_free} }  
\end{figure}

\newpage

\section*{Singlet--triplet splittings of intermediate \textbf{3}}

\begin{table*}[htb!]
\centering
\caption{Singlet--triplet ($\Delta E=E_T - E_S$) splitting for \textbf{3} in kJ/mol in the present and in a previous study (obtained with TPSS-D3/def2-SV(P) without His147 in the QM region).  
In addition, the QM/MM energy components from Eqn. 1 are also included (with $\Delta E_{\text{MM}} = \Delta E_{\text{MM123}} - \Delta E_{\text{MM1}}$), as well as  
the QM+ptch energy calculated with the larger def2-TZVPP basis set  (single-point energy calculation; $\Delta E_{\text{QM}}$). \label{single_triplet_splittings} }
\begin{tabular}{lllcccc}
 \hline\hline \\[-2.0ex]
 Source  &     System 2                    &  Func    & $\Delta E_{\text{QM/MM}}$ & $ \Delta E_{\text{QM+ptch}}$ & $\Delta E_{\text{QM}}$ & $\Delta E_{\text{MM}}$  \\[0.5ex]
\hline \\[-1.5ex] 
QM/MM \cite{hedegaard2017} & free  & TPSS-D3  & 12.5                      & 10.6                         &  14.0                 &  2.0	                   \\[0.5ex]
QM/MM \cite{hedegaard2017} & free  & B3LYP-D3 & 13.2                      & 11.2                         &  16.5                 &  2.0                     \\[0.5ex] 
QM/MM                      & free  & TPSS-D3  & 11.9                      & 10.3                         &  14.2                 &  1.5	                   \\[0.5ex]
QM/MM                      & free  & B3LYP-D3 & 14.0                      & 12.4                         &  17.0              &  1.5                     \\[0.5ex]
\hline \hline  
\end{tabular} 
\end{table*}

\newpage

\section*{Reaction profiles and energetics for intermediates \textbf{3}--\textbf{7}}

\begin{table}[htb!]
\centering
\begin{tabular}{lcccc}
    \hline \hline \\[-2.0ex]
    Reaction        &  \multicolumn{4}{c}{\textbf{3}$\rightarrow$\textbf{4a}}       \\[0.5ex] 
        \hline \\[-1.5ex]
      $\Delta E$    &  \multicolumn{2}{c}{TPSS-D3} & \multicolumn{2}{c}{B3LYP-D3}  \\[0.5ex]
    Struct                    &   TS &  Prod      &  TS       & Prod   \\[0.5ex] 
    $\Delta E_{\text{QM/MM}}$  &  14.9    & -5.6  & 17.0    & 0.1   \\[0.5ex] 
    $\Delta E_{\text{QM+ptch}}$&  17.8    & -3.6  & 19.9    & 2.0   \\[0.5ex] 
    $\Delta E_{\text{MM}}$     &  -2.9    & -1.9  & -2.9    & -1.9  \\[0.5ex] 
    $\Delta E_{\text{QM}}$     &  24.5    &  52.1 & 27.2    & 63.9  \\[0.5ex]
    $\Delta E_{\text{ptch}}$   &  -6.7    & -55.7 & -7.3    & -61.8 \\[0.5ex]
    \hline \hline
   \end{tabular}
\caption{Reaction energies and barriers (kJ/mol) obtained from single-point calculations with the def2-TZVPP basis set (for the meaning of the different energies, see Models and Methods section of the main paper).  
\label{cu-o2_his_prot_tab}}
\end{table}
\begin{figure}[htb!]
\centering
\includegraphics[width=0.75\textwidth]{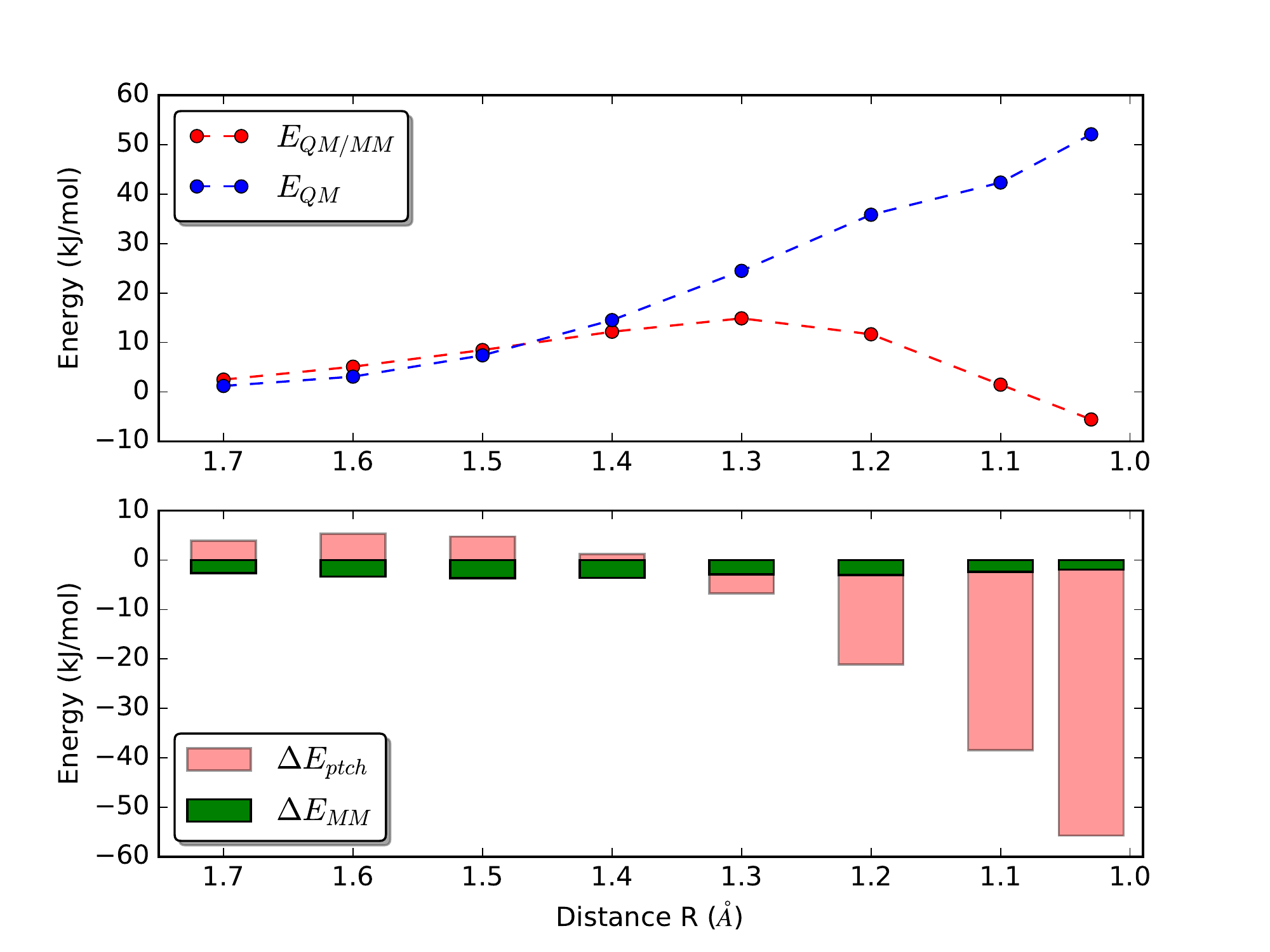}  
\caption{$\Delta E_{\text{QM/MM}}$ and $\Delta E_{\text{QM}}$ for a range of \ce{O-H} distances (reaction \textbf{3}$\rightarrow$\textbf{4a}). In the lower figure, the electrostatic and MM contributions to $\Delta E_{\text{QM/MM}}$ are shown for the various distances.  \label{cu-o2_his_prot_fig}} 
\end{figure}

\begin{figure}[htb!]
\centering
\includegraphics[width=0.48\textwidth]{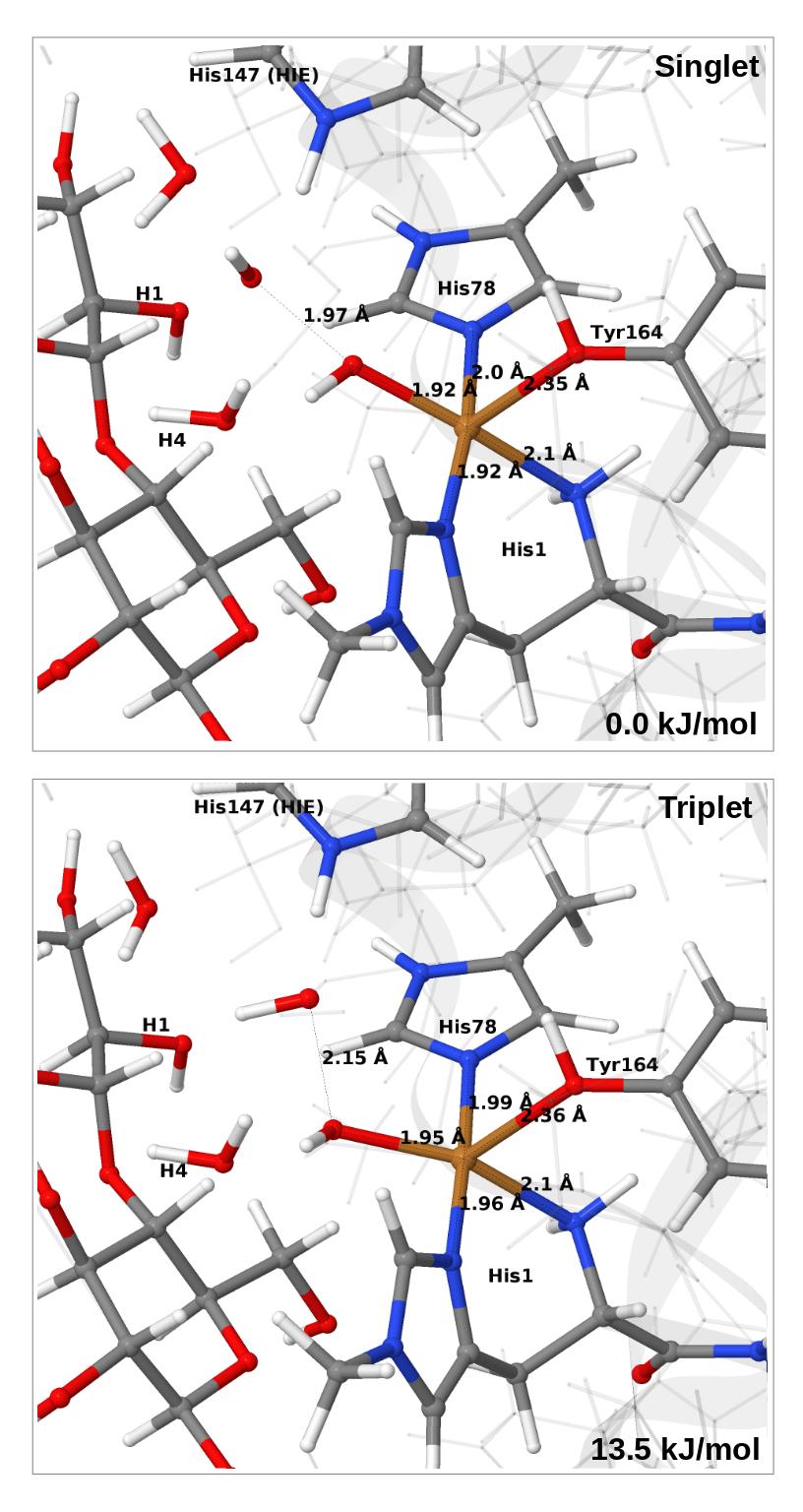} 
\caption{Optimized structures of state \textbf{5} with HIE147 in the singlet (\textit{top}) and triplet (\textit{bottom}) spin-states. \label{cu-ohoh_hip147_coordinated}}
\end{figure}

\begin{table}[htb!]
\centering
\small
\begin{tabular}{lcccccccc}
    \hline \hline \\[-2.0ex] 
    Reaction        &  \multicolumn{4}{c}{\textbf{4b}$\rightarrow$\textbf{6a}}       \\[0.5ex] 
    \hline \\[-1.5ex]
      $\Delta E$    &  \multicolumn{2}{c}{TPSS-D3} & \multicolumn{2}{c}{B3LYP-D3}  \\[0.5ex]
      Energy                     & TS             &  Prod         &  TS         & Prod        \\[0.5ex] 
    $\Delta E_{\text{QM/MM}}$    &    47.8        &  -0.5         &  66.2       &   54.1      \\[0.5ex] 
    $\Delta E_{\text{QM+ptch}}$  &    54.7        &   1.7         &  68.5       &   56.3      \\[0.5ex] 
    $\Delta E_{\text{MM}}$       &    -6.9        &  -2.2        &  -6.9      &   -2.2     \\[0.5ex] 
    $\Delta E_{\text{QM}}$       &   110.5        &  98.2         &   165.6     &   157.8      \\[0.5ex]
    $\Delta E_{\text{ptch}}$     &   -55.8        & -96.4         &  -97.1     &   -101.5     \\[0.5ex]
   \hline \hline
   \end{tabular}    
\caption{Reaction energies and barriers (kJ/mol) obtained from single-point calculations with the def2-TZVPP basis set (for the meaning of the different energies, see Models and Methods section of the main paper).  \label{cu-ooh_his_prot_h2o_diss_table}}
\end{table}
\begin{figure}[ht!]
\centering
\includegraphics[width=0.75\textwidth]{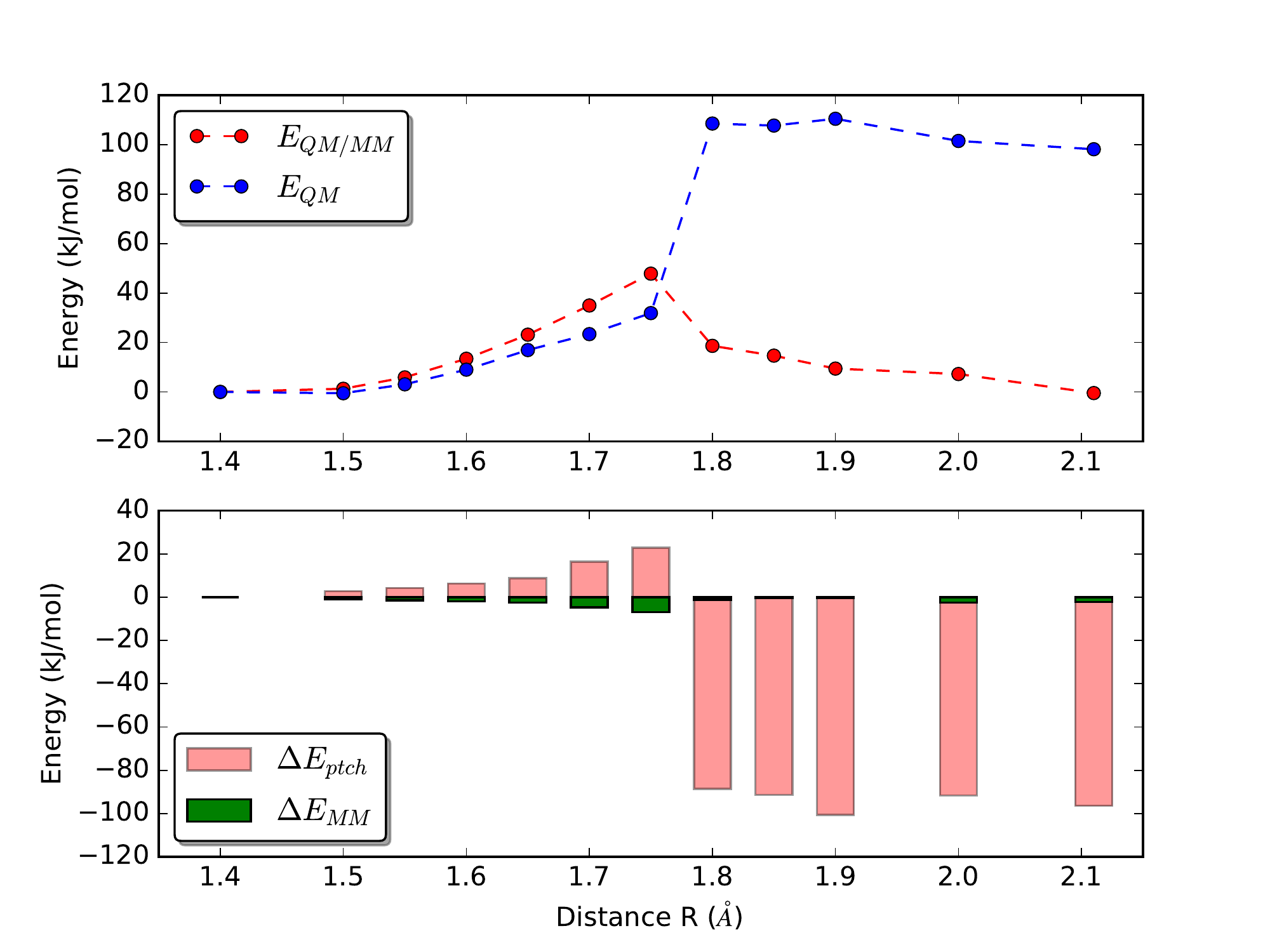}  
\caption{$\Delta E_{\text{QM/MM}}$ and $\Delta E_{\text{QM}}$ for a range of \ce{O-O} distances in the water dissociation reaction \textbf{4b}$\rightarrow$\textbf{6a}. In the lower figure, the electrostatic and MM contributions to $\Delta E_{\text{QM/MM}}$ are shown for the various distances.  \label{cu-ooh_red_his_prot}}
\end{figure}

\begin{table}[ht!]
\centering
\small
\begin{tabular}{lcccccccc}
    \hline \hline \\[-2.0ex]
    Reaction        &  \multicolumn{4}{c}{\ce{C4-H} abstraction by \textbf{3} 	}       \\[0.5ex] 
    \hline \\[-1.5ex]
      $\Delta E$    &  \multicolumn{2}{c}{TPSS-D3} & \multicolumn{2}{c}{B3LYP-D3}  \\[0.5ex]
       Method   & \multicolumn{2}{c}{TPSS-D3} & \multicolumn{2}{c}{B3LYP-D3} &  \\[0.5ex] 
    $\Delta E_{\text{QM/MM}}$   &   155.7          &   140.3       &  168.3    &  128.0      \\[0.5ex] 
    $\Delta E_{\text{QM+ptch}}$ &   154.5          &   135.6       &  167.1    &  123.2      \\[0.5ex] 
    $\Delta E_{\text{MM}}$      &     1.1          &     4.7       &    1.1    &  4.7        \\[0.5ex] 
    $\Delta E_{\text{QM}}$      &   151.5          &   133.5       &  162.2    &  123.6      \\[0.5ex]
    $\Delta E_{\text{ptch}}$    &     3.0          &     2.1       &    4.9   &  -0.4       \\[0.5ex]
   \hline \hline
   \end{tabular}
\caption{Reaction energies and barriers obtained with def2-TZVPP singlet-point calculations. His147 was modelled as HIE. \label{cu-o2_rh_reaction}}
\end{table}
\begin{figure}[tbh!]
\centering
\includegraphics[width=0.75\textwidth]{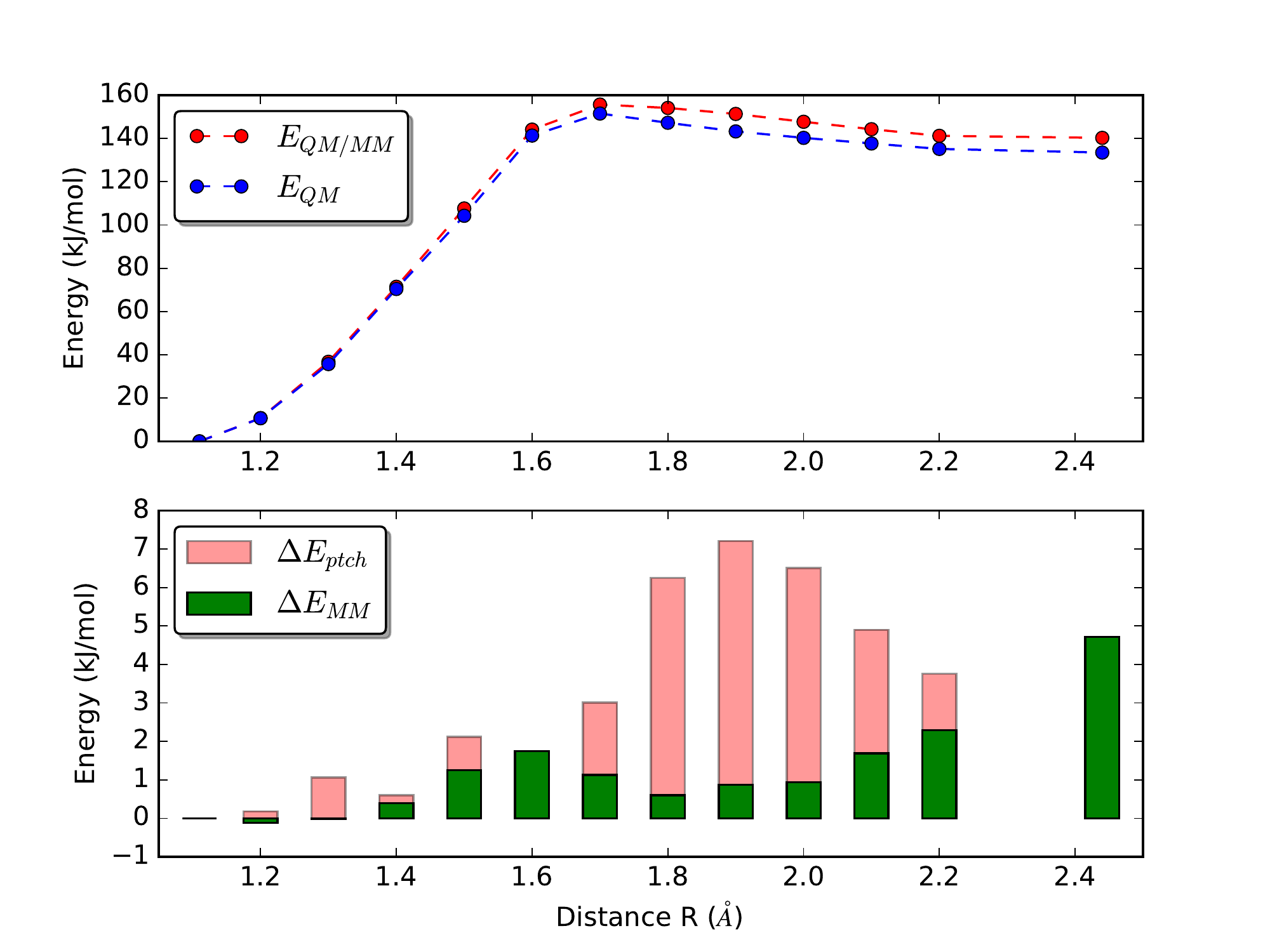}  
\caption{Reaction energies and barrier for \ce{C-H} hydrogen abstraction by \ce{[CuO2]+}. The figure  shows $\Delta E_{\text{QM/MM}}$ and $\Delta E_{\text{QM}}$ for a range of \ce{O-H} distances. In the lower figure, the electrostatic and MM contributions to $\Delta E_{\text{QM/MM}}$ are shown for the various distances.  \label{cu-o2_his_h-transfer}}
\end{figure}

\begin{table}[ht!]
\centering
\small
\begin{tabular}{lcccccccc}
    \hline \hline \\[-2.0ex]
    Reaction        &  \multicolumn{4}{c}{\textbf{6c}$\rightarrow$\textbf{7b} (HID147)} & \multicolumn{4}{c}{\textbf{6b}$\rightarrow$\textbf{7a}  (HID147)}       \\[0.5ex] 
    \hline \\[-1.5ex]
      $\Delta E$    & \multicolumn{2}{c}{TPSS-D3} & \multicolumn{2}{c}{B3LYP-D3} & \multicolumn{2}{c}{TPSS-D3} & \multicolumn{2}{c}{B3LYP-D3}  \\[0.5ex]
                                &   TS           &  Prod   &  TS    & Prod    &   TS     &  Prod       &  TS    & Prod     \\[0.5ex]   
    $\Delta E_{\text{QM/MM}}$   &  93.8          &  48.5   &  97.0   & -22.2   &   68.5   &  -17.2      &  72.9  &  -21.8   \\[0.5ex] 
    $\Delta E_{\text{QM+ptch}}$ &  75.8          &  31.4   &  84.2  & -39.4   &   67.0   &  -23.7      &  71.3  &  -28.4   \\[0.5ex] 
    $\Delta E_{\text{MM}}$      &  12.8          &  17.2   &  12.8  &  17.2   &   1.6    &   6.6       &   1.6     &   6.6    \\[0.5ex] 
    $\Delta E_{\text{QM}}$      &  76.7          &  44.8   &  82.8  &  -3.9   &   68.2   &   -12.3     &   73.9 &  -16.8   \\[0.5ex]
    $\Delta E_{\text{ptch}}$    &  -2.8          &  -13.5  &  1.4   &  -35.5  &   -1.2   &   -11.4     &  -2.6  &  -11.6  \\[0.5ex]
   \hline \hline
   \end{tabular}     
\caption{Reaction energies and barriers for the hydrogen abstraction from \ce{RH} by the \ce{[CuO]+} (\textbf{6b}) and \ce{[CuOH]^{2+}} (\textbf{6c}) states. The energies are obtained with def2-TZVPP singlet-point calculations.  \label{cu-oh_and_cuo_rh_reactions_hid}}
\end{table}
\begin{figure}[tbh!]
\centering
\includegraphics[width=0.75\textwidth]{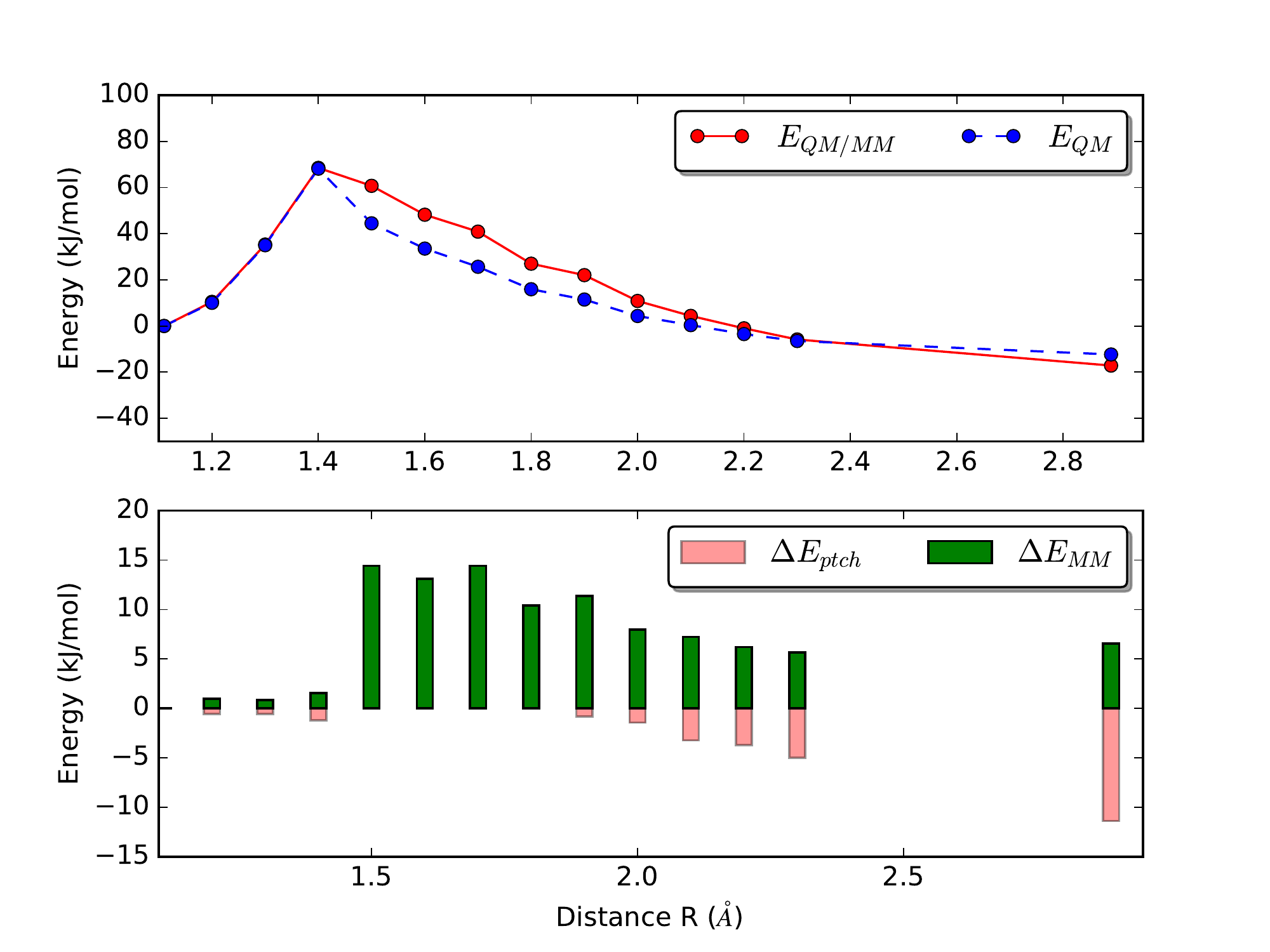}  
\caption{Reaction energies and barriers for \ce{C-H} hydrogen abstraction by \ce{[CuO]+} (\textbf{6b}). The figure  shows $\Delta E_{\text{QM/MM}}$ and $\Delta E_{\text{QM}}$ for a range of \ce{C4-H} distances.  In the lower figure, the electrostatic and MM contributions to $\Delta E_{\text{QM/MM}}$ are shown for the various distances.  \label{cu-o_his_h-transfer}}
\end{figure}
\begin{table}[ht!]
\centering
\small
\begin{tabular}{lcccccccc}
    \hline \hline \\[-2.0ex]
    Reaction        &  \multicolumn{4}{c}{\textbf{6c}$\rightarrow$\textbf{7b}  (HIE147)} & \multicolumn{4}{c}{\textbf{6b}$\rightarrow$\textbf{7a}  (HIE147)}       \\[0.5ex] 
    \hline \\[-1.5ex]
      $\Delta E$    & \multicolumn{2}{c}{TPSS-D3} & \multicolumn{2}{c}{B3LYP-D3} & \multicolumn{2}{c}{TPSS-D3} & \multicolumn{2}{c}{B3LYP-D3}  \\[0.5ex]
                                &   TS           &  Prod   &  TS    & Prod    &   TS     &  Prod       &  TS          & Prod     \\[0.5ex]   
    $\Delta E_{\text{QM/MM}}$   &     103.4      &   56.7  &  92.9  &   -13.5 &   104.5  &  -22.4      &  111.1       & -27.1	 \\[0.5ex] 
    $\Delta E_{\text{QM+ptch}}$ &     85.5       &   39.9  &  92.9  &   -30.3 &   102.4  &  -25.4      &  109.0       & -30.1	 \\[0.5ex] 
    $\Delta E_{\text{MM}}$      &     17.9       &   16.8  &  0.0   &    16.8 &     2.2  &   3.0       &  2.2         &  3.0     \\[0.5ex] 
    $\Delta E_{\text{QM}}$      &     94.2       &   59.4  &  -     &   -14.8 &   100.9  &  -20.1      &  112.3       & -24.8	 \\[0.5ex]
    $\Delta E_{\text{ptch}}$    &     -8.7       &   -19.5 &  -     &   -15.6 &     1.5  &  -5.4       &  -3.3       & -5.3	 \\[0.5ex]
   \hline \hline
   \end{tabular}     
\caption{Reaction energies and barriers for hydrogen abstraction from \ce{RH} by the \ce{[CuO]+} (\textbf{6b}) and \ce{[CuOH]^{2+}} (\textbf{6c}) moieties. The energies are obtained with def2-TZVPP singlet-point calculations.  \label{cu-oh_and_cuo_rh_reactions_hie}}
\end{table}
\begin{figure}[tbh!]
\centering
\includegraphics[width=0.75\textwidth]{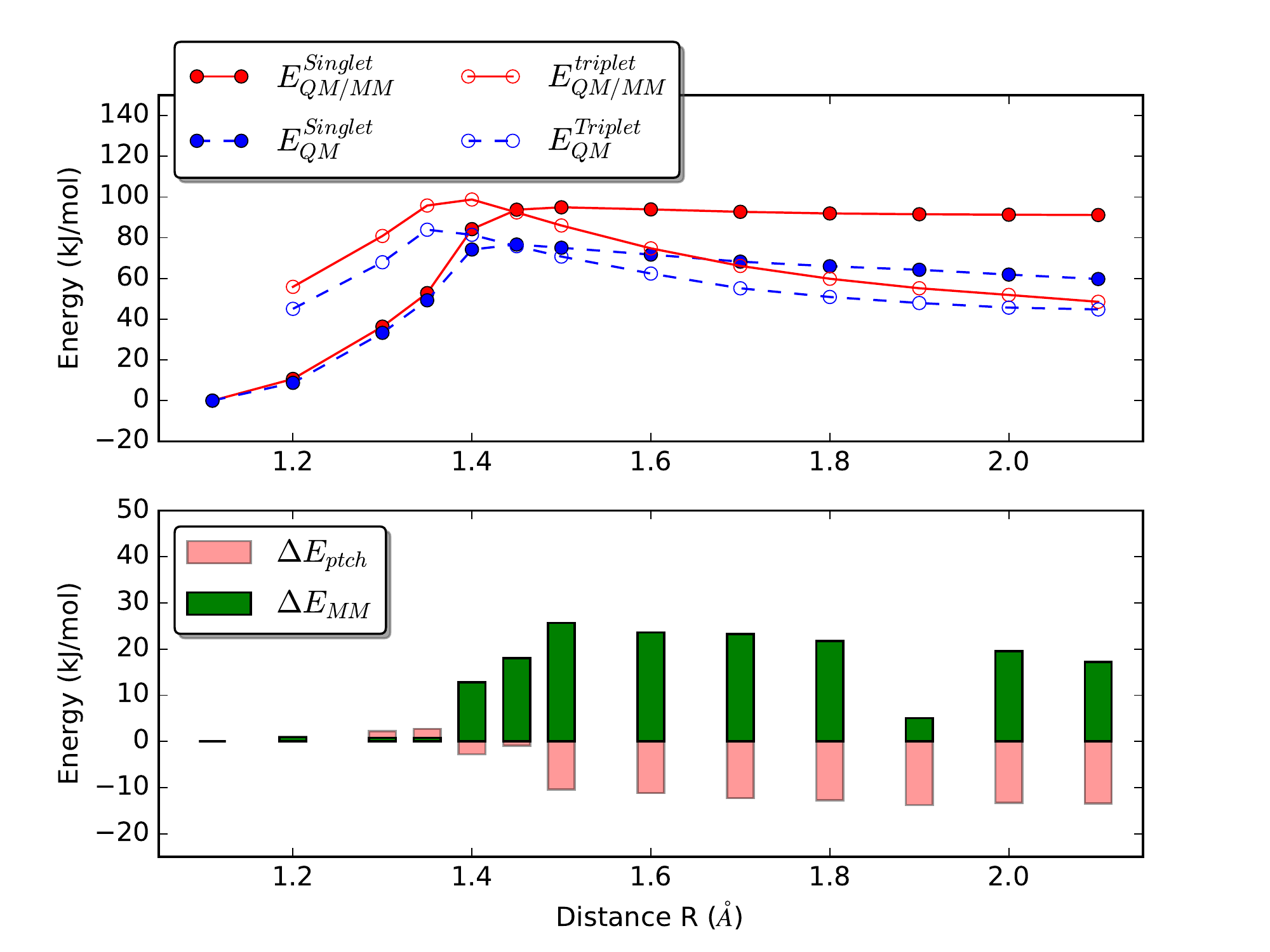}  
\caption{Reaction energies and barriers for \ce{C-H} hydrogen abstraction by \ce{[CuOH]^{2+}} (\textbf{6c}). The figure  shows $\Delta E_{\text{QM/MM}}$ and $\Delta E_{\text{QM}}$ for a range of \ce{O-H} distances.  In the lower figure, the electrostatic and MM contributions to $\Delta E_{\text{QM/MM}}$ are shown for the various distances (always for the most stable state, singlet or triplet).  \label{cu-oh_his_h-transfer}}
\end{figure} 

\begin{table}[htb!]
\centering
\small
\begin{tabular}{lcccccccc}
    \hline \hline \\[-2.0ex] 
	Reaction        &  \multicolumn{4}{c}{\textbf{7a}$\rightarrow$\textbf{8a} (HIE)}  &  \multicolumn{4}{c}{\textbf{7a}$\rightarrow$\textbf{8a} (HID)}       \\[0.5ex] 
    \hline \\[-1.5ex]
      $\Delta E$    &  \multicolumn{2}{c}{TPSS-D3} & \multicolumn{2}{c}{B3LYP-D3} & \multicolumn{2}{c}{TPSS-D3} & \multicolumn{2}{c}{B3LYP-D3}  \\[0.5ex]
      Energy                     & TS             &  Prod         &  TS         & Prod       & TS        &  Prod    &  TS    & Prod       \\[0.5ex] 
    $\Delta E_{\text{QM/MM}}$    &  40.3          &  -182.6       &  53.4       & -193.1     & 35.1 &  -194.0  &  44.3  &  -203.8  \\[0.5ex] 
    $\Delta E_{\text{QM+ptch}}$  &  30.0          &  -168.6       &  31.5       & -205.1     &  23.2 &  -218.5  &  22.9  &  -228.3 \\[0.5ex] 
    $\Delta E_{\text{MM}}$       &  10.3          &   12.0        &  21.9       &   12.0     & 11.9 &    24.5  &  21.4  &    24.5  \\[0.5ex] 
    $\Delta E_{\text{QM}}$       &  38.8          &  -154.3       &  36.6       & -190.2     & 30.1 &  -203.5  &  21.2  &  -213.0  \\[0.5ex]
    $\Delta E_{\text{ptch}}$     &  -8.8          &    -14.4      &  -5.2       & -15.0      &  -6.9 &   -14.4  &   1.8  &  -15.3 \\[0.5ex]
   \hline \hline
   \end{tabular}    
\caption{Reaction energies and barriers (kJ/mol) for the recombination step (\textbf{7a}$\rightarrow$\textbf{8a}) with His147 in either the HIE or the HID state. The energies are obtained from single-point calculations with the def2-TZVPP basis set. \label{cu-oh_recom_hie}}
\end{table}

\begin{figure}[tbh!]
\centering
\includegraphics[width=0.75\textwidth]{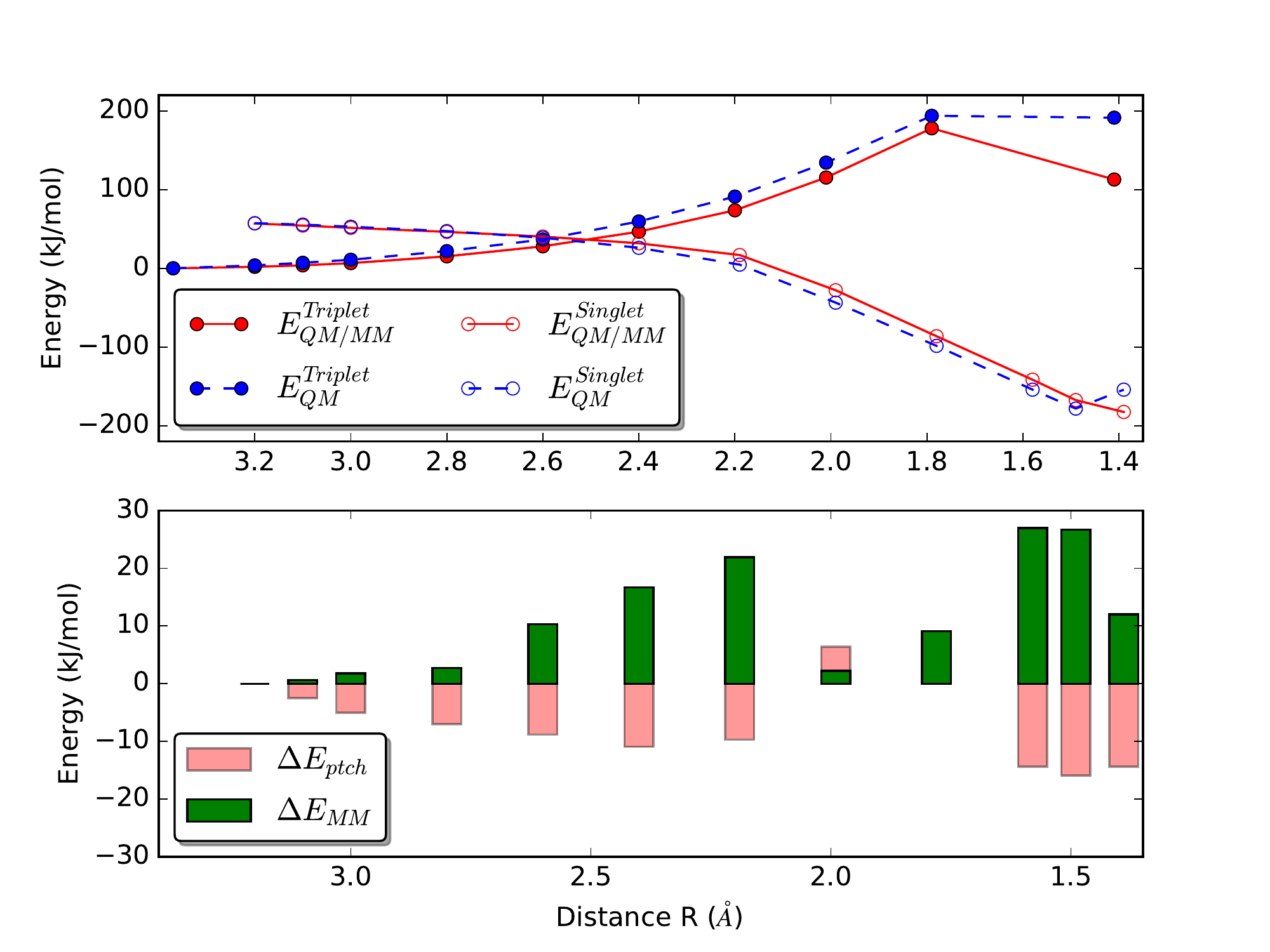}  
\caption{Reaction energies and barrier for the recombination step (\textbf{7a}$\rightarrow$\textbf{8a}) with His147 in the HIE state. The figure  shows $\Delta E_{\text{QM/MM}}$ and $\Delta E_{\text{QM}}$ for a range of \ce{C-O} distances.  In the lower figure, the electrostatic and MM contributions to $\Delta E_{\text{QM/MM}}$ are shown for the various distances.  \label{cu-oh_recom_hie_not_prot}}
\end{figure}
\begin{table}[htb!]
\centering
\small
\begin{tabular}{lcccccccc}
    \hline \hline \\[-2.0ex] 
	Reaction        &  \multicolumn{4}{c}{\textbf{7a}$\rightarrow$\textbf{8a} (HIP)}       \\[0.5ex] 
    \hline \\[-1.5ex]
      $\Delta E$    &  \multicolumn{2}{c}{TPSS-D3} & \multicolumn{2}{c}{B3LYP-D3}  \\[0.5ex]
      Energy                     & TS              &  Prod         &  TS         & Prod        \\[0.5ex] 
    $\Delta E_{\text{QM/MM}}$    &  61.9           &  -158.1       &  53.1       &  -165.7     \\[0.5ex] 
    $\Delta E_{\text{QM+ptch}}$  &  51.5           &  -165.7       &  48.7       &  -156.8     \\[0.5ex] 
    $\Delta E_{\text{MM}}$       &  47.5           &  -156.8       &  4.4        &  -163.1     \\[0.5ex] 
    $\Delta E_{\text{QM}}$       &  10.5           &     7.6       &   28.1      &     7.6     \\[0.5ex]
    $\Delta E_{\text{ptch}}$     &  3.9            &    -8.8       &   20.5      &    -7.0     \\[0.5ex]
   \hline \hline
   \end{tabular}    
\caption{Reaction energies and barriers (kJ/mol) obtained for the recombination step (\textbf{7a}$\rightarrow$\textbf{8a}) with His147 in the HIP state. The energies are obtained from single-point calculations with the def2-TZVPP basis set . \label{cu-oh_recom_hip}}
\end{table}

\newpage

\section*{}
\bibliography{lpmo}